\documentstyle[psfig,mncite]{mn}

\newif\ifAMStwofonts

\newcommand{\abell}{{\it Abell\,370}}
\newcommand{\abs}{absorption line}

\newcommand{\cl}{{\it CL\,0949$+$44}}
\newcommand{\dmgb}{\langle\Delta \mbox{Mg}_b\rangle}
\newcommand{\dmb}{\langle\Delta \mbox{M}_B\rangle}
\newcommand\egs{elliptical galaxies}
\newcommand\es{ellipticals}
\newcommand{\etg}{early--type galaxies}
\newcommand{\fef}{Fe\,{\small 5270}}
\newcommand{\fes}{Fe\,{\small 5335}}
\newcommand{\fjr}{Faber--Jackson relation}
\newcommand{\hauf}{galaxy cluster}
\newcommand{\hb}{H$_\beta$}
\newcommand{\hd}{H$_\delta$}
\newcommand{\lst}{linestrength}
\newcommand{\hst}{{\it Hubble} space telescope}
\newcommand{\lo}{$\lambda_0 = $}
\newcommand{\loap}{$\lambda_0 \approx $}
\newcommand\lum{luminosity}

\newcommand{\mdo}{\,\rmn mag}
\newcommand{\met}{metallicity}
\newcommand{\mg}{Mg$_b$}
\newcommand{\mga}{Mg$_b$ ab\-sorp\-tion line}
\newcommand{\mgi}{Mg$_b$ index}
\newcommand{\mgl}{Mg$_b$ linestrength}
\newcommand{\mgs}{Mg$_b$--$\sigma$}
\newcommand{\bmgs}{Mg$_{\bmath{b}}$--$\bsigma$}
\newcommand{\mgsr}{Mg$_b$--$\sigma$ relation}
\newcommand{\mgst}{Mg$_b$--$\sigma$ test}
\newcommand{\mgzsr}{Mg$_2$--$\sigma$ relation}

\newcommand{\ms}{{\it MS\,1512$+$36}}
\newcommand{\odrei}{{\tiny $\triangle$}}
\newcommand{\oquad}{{\tiny $\Box$}}
\newcommand{\phot}{photometric}
\newcommand{\pop}{population synthesis}
\newcommand{\qo}{deceleration pa\-ra\-me\-ter}
\newcommand{\rsr}{$r_e$--$\sigma$ relation}
\newcommand{\rv}{redshift}
\newcommand{\snr}{signal--to--noise ratio}
\newcommand{\spec}{spectroscopic}
\newcommand\veldis{velocity dispersion}

\ifoldfss
  \newcommand{\rmn}[1] {{\rm #1}}

  \ifCUPmtlplainloaded \else
    \NewTextAlphabet{textbfit} {cmbxti10} {}
    \NewTextAlphabet{textbfss} {cmssbx10} {}
    \NewMathAlphabet{mathbfit} {cmbxti10} {} 
    \NewMathAlphabet{mathbfss} {cmssbx10} {} 
  \fi
  \ifAMStwofonts
    \ifCUPmtlplainloaded \else
      \NewSymbolFont{upmath} {eurm10}
      \NewSymbolFont{AMSa} {msam10}
      \NewMathSymbol{\upi}     {0}{upmath}{19}
      \NewMathSymbol{\umu}     {0}{upmath}{16}
      \NewMathSymbol{\upartial}{0}{upmath}{40}
      \NewMathSymbol{\leqslant}{3}{AMSa}{36}
      \NewMathSymbol{\geqslant}{3}{AMSa}{3E}

       \let\le=\leqslant
       \let\ge=\geqslant
    \fi
  \fi
\fi 

\ifnfssone
  \newmathalphabet{\mathit}
  \addtoversion{normal}{\mathit}{cmr}{m}{it}
  \addtoversion{bold}{\mathit}{cmr}{bx}{it}
  \newcommand{\rmn}[1] {\mathrm{#1}}

  \newmathalphabet{\mathbfit} 
  \addtoversion{normal}{\mathbfit}{cmr}{bx}{it}
  \addtoversion{bold}{\mathbfit}{cmr}{bx}{it}
  \newmathalphabet{\mathbfss} 
  \addtoversion{normal}{\mathbfss}{cmss}{bx}{n}
  \addtoversion{bold}{\mathbfss}{cmss}{bx}{n}
  \ifAMStwofonts
    \ifCUPmtlplainloaded \else
      \UseAMStwoboldmath
      \makeatletter
      \new@mathgroup\upmath@group
      \define@mathgroup\mv@normal\upmath@group{eur}{m}{n}
      \define@mathgroup\mv@bold\upmath@group{eur}{b}{n}
      \edef\UPM{\hexnumber\upmath@group}
      \new@mathgroup\amsa@group
      \define@mathgroup\mv@normal\amsa@group{msa}{m}{n}
      \define@mathgroup\mv@bold\amsa@group{msa}{m}{n}
      \edef\AMSa{\hexnumber\amsa@group}
      \makeatother
      \mathchardef\upi="0\UPM19
      \mathchardef\umu="0\UPM16
      \mathchardef\upartial="0\UPM40
      \mathchardef\leqslant="3\AMSa36
      \mathchardef\geqslant="3\AMSa3E

       \let\le=\leqslant
       \let\ge=\geqslant
    \fi
  \fi
\fi 

\ifnfsstwo
  \newcommand{\rmn}[1] {\mathrm{#1}}

  \DeclareMathAlphabet{\mathbfit}{OT1}{cmr}{bx}{it}
  \SetMathAlphabet\mathbfit{bold}{OT1}{cmr}{bx}{it}
  \DeclareMathAlphabet{\mathbfss}{OT1}{cmss}{bx}{n}
  \SetMathAlphabet\mathbfss{bold}{OT1}{cmss}{bx}{n}
  \ifAMStwofonts
    \ifCUPmtlplainloaded \else
      \DeclareSymbolFont{UPM}{U}{eur}{m}{n}
      \SetSymbolFont{UPM}{bold}{U}{eur}{b}{n}
      \DeclareSymbolFont{AMSa}{U}{msa}{m}{n}
      \DeclareMathSymbol{\upi}{0}{UPM}{"19}
      \DeclareMathSymbol{\umu}{0}{UPM}{"16}
      \DeclareMathSymbol{\upartial}{0}{UPM}{"40}
      \DeclareMathSymbol{\leqslant}{3}{AMSa}{"36}
      \DeclareMathSymbol{\geqslant}{3}{AMSa}{"3E}

       \let\le=\leqslant
       \let\ge=\geqslant
    \fi
  \fi
\fi 

\ifCUPmtlplainloaded \else
  \ifAMStwofonts \else 
    \def\upi{\pi}
    \def\umu{\mu}
    \def\upartial{\partial}
  \fi
\fi

\title[The Mg$_b$--$\sigma$ Relation at $z\approx0.37$]
{The Mg$_{\bmath{b}}$--$\bsigma$ Relation of Elliptical Galaxies at 
$\bmath{z\approx0.37}$
\thanks{Partly based on observations carried out at the
European Southern Observatory, La Silla, Chile.}}
\author[B. L. Ziegler and R. Bender]
       {Bodo L. Ziegler\thanks{Visiting astronomer of the German--Spanish
Astronomical Center, Calar Alto, operated by the Max--Planck--Institut f\"ur
Astronomie, Heidelberg, jointly with the Spanish National Commission for
Astronomy.}\thanks{E-mail: ziegler@usm.uni-muenchen.de} and Ralf Bender
\\
Institut f\"ur Astronomie und Astrophysik der 
Ludwigs--Maximilian--Universit\"at, Universit\"atssternwarte, 
Scheinerstra\ss e 1, \\ 81679~M\"unchen, Germany}
\date{Accepted May 1997 .
      Received April 1997 ;
      in original form February 1997 }

\pagerange{\pageref{firstpage}--\pageref{lastpage}}
\pubyear{1997}

\begin{document}

\maketitle

\label{firstpage}

\begin{abstract}
We derive \abs\ indices of \egs\ in clusters at intermediate \rv\
($z\approx0.37$) from medium-resolution spectroscopy together
with kinematical parameters. These galaxies exhibit a relationship
between the \lst\ of \mg\ (\loap\ 5170\,\AA) and their internal \veldis\
$\sigma$ similar to local dynamically hot galaxies. But for any given $\sigma$,
the \mgl\ of the distant ellipticals is significantly lower than the mean
value of the nearby sample. The difference of \mg\ between the two samples
is small ($\dmgb \,\approx -0.4$\,\AA) and can be fully
attributed to the younger age of the distant stellar populations in
accordance with the passive evolution model for \egs. The low reduction of
\mg\ at a look--back time of about 5 Gyrs requires that the bulk of the stars
in cluster \es\ have formed at very high redshifts of $z_f>2$. For the most
massive galaxies, where the reduction is even lower, $z_f$ probably exceeds 4.

Unlike most methods to measure the evolution of \egs\ using luminosities,
surface brightnesses or colours, the \mgst\ does not depend on corrections
for extinction and cosmic expansion (K--correction) and only very little on
the slope of the initial mass function. The combination of a kinematical
parameter with a stellar population indicator allows us to study the evolution
of very similar objects. In addition, the good mass estimate provided
by $\sigma$ means that the selection criteria for the galaxy sample as a whole 
are well controlled.

In quantitative agreement with the reduction of the \mg\ absorption we find
an increase of the $B$ magnitude of $\dmb \,\approx -0.5\mdo$
at fixed $\sigma$ from the \fjr. The brightening of the \es\ at $z=0.37$
arises solely from the evolution of their stellar populations and is of the
same order as the change in magnitudes when varying the \qo\ $q_0$ from
$-0.5$ to $+0.5$ at this redshift.

Studying the evolution of the \mgsr\ in combination with that of the \fjr\
allows us to constrain both the slope of the initial mass function and the
value of the \qo. Our current data with their measurement errors are
compatible with the standard Salpeter IMF and $q_0=0.5\pm 0.5$.
\end{abstract}


\section{Introduction}

Twenty years after the seminal papers on the formation of \egs\ by 
\scite{Larso75} and by 
\scite{Toomr77} there is still much
disagreement among the astronomical community on both the process of
formation and the evolution of early-type galaxies. Is an elliptical galaxy
formed in a single collapse or via merging? Was there a short epoch of
formation or have \es\ been formed continuously by hierchical merging at
similar levels? What is the influence of the density environment? Once
created, is the stellar population of \es\ evolving just passively or do
minor merging/accretion events drastically change their characteristics
frequently?

In the local universe, ongoing merging is observed and it is generally
assumed that most of the `ultra-luminous' {\it IRAS} galaxies represent
merging processes \cite{Schwe90}. Numerical simulations show in great detail
how the merging of two spiral galaxies leads to the formation of a stellar
system with a de~Vaucouleurs profile \cite{BH92}. Often, a core
kinematically decoupled from the main body is found in the interior of such
simulated merger products implying that most of the ellipticals observed to
have a decoupled core were formed in a merger. More generally, \egs\ with
boxy or irregular isophotes are thought to be the result of
mergers/interactions \cite{BSDMM89} indicating that at least 2/3 of all
bright \es\ have a merger origin.

Morphological and \spec\ examinations of \hauf s at intermediate \rv s have
shown that the `Butcher--Oemler effect' is not due to a significantly
increasing merger rate but to an increasing star formation activity of disk
galaxies with redshift (e.g. \pcite{DOBG94}). Rather, the very low scatter
in the optical/infrared colour--magnitude diagrams \cite{BLE92} and the
\mgzsr\ \cite{BBF93} of massive \egs\ in nearby clusters is compatible with
a short formation epoch at a high redshift ($z_f>2$) implying that recent
mergers add only a very small fraction to the total number of cluster \es.
Indeed, stellar \pop\ models can fit best the spectral energy distribution
of observed local \es\ assuming a short period of star formation followed by
`passive' evolution of the stellar population without any significant new
star formation at later times (e.g. \pcite{BC93}). Recently, accurate
measurements of physical relations of early-type galaxies at intermediate
\rv s ($z\approx0.4$) support this passive evolution scenario. Thus, the
Tolman test does not indicate a significant deviation of the surface
brightness evolution from the pure cosmological dependence (e.g.
\pcite{PDC96}, \pcite{SBL97}). Analysis of the fundamental plane yield a
moderate decrease in the $M/L$ ratio (e.g. \pcite{DF96}, \pcite{KDFIF97}).
We presented preliminary results from an investigation of the \mgsr\ showing
mild evolution of the \mgi\ and the blue luminosity \cite{BZB96}. In
addition, observational data out to $z\approx 1$ are accumulating that are
compatible with this `passive evolution' scenario for \es\ (or rather, the
more luminous, red galaxies at higher \rv s) both in the field and in
clusters. The various methods to test evolution comprise luminosity
functions and number counts (e.g. \pcite{GPMC95},
\pcite{CFRS6}, \pcite{ECBHG95}), optical and near infrared
colour--magnitude diagrams (e.g. \pcite{AECC93}, \pcite{SED95}) and
projections of the fundamental plane relations (e.g. \pcite{SCYLE96}).
Taking advantage of the capability of the \hst\ and 10m-telescopes, the
search for galaxies at high \rv\ has just begun using both morphological and
\spec\ information. E.g., \scite{SGPDA96} have found candidate precursors of
\es\ at $3<z<3.5$.

Semi-analytic models of galaxy formation based on CDM-like structure
formation theory have shown that \egs\ could have been formed in mergers and
nevertheless appear so homogeneous in their stellar population as is
observed in the local Universe \cite{Kauff96}. In these models, most of
the stars in ellipticals formed at high redshifts ($z_f>1.9$).

But, is the `passive evolution' scenario valid for the whole population of
nearby \etg\ or could it be that the methods stated above pick up only those
galaxies that comply with the assumptions of passive evolution? Using the 
$V/V_{max}$--test \cite{Schmi68}, \scite{KCW97} find that the
fraction of early-type galaxies dropping out of their sample increases with
\rv\ so that at $z\approx 1$ only about one third of the bright E and S0
galaxies seen today were already assembled. Since this investigation is
based on fields of the Canada--France redshift survey that contains mostly
field galaxies, their result may indicate more rapid number density
evolution in the low density environment. CDM simulations by \scite{BCF97}
predict that a galaxy may change its appearance as disk-like or spheroidal
several times during its existence. Infrared observations of M32
(\pcite{ES92}, \pcite{Freed92}) and the bulge of M31 \cite{RM91} resolve a
population of very bright red giant stars that indicate a generation of
stars only about 5~Gyrs old.

In this paper, we present a new method to examine the evolution of the
stellar population of \egs\ with redshift based on the tight relationship
between the \mgi\ and the velocity dispersion $\sigma$ of \egs. This method
permits good control of the sample selection and has several advantages with
respect to those mentioned above. It will be described in Section 2, wheras
Section 3, 4 and 5 will deal with the sample selection, the observations and
the data reduction, respectively. Our results and conclusions will be
presented in Sections 6 and 7.

\section[]{The \bmgs\ test}

\begin{figure*}
\psfig{figure=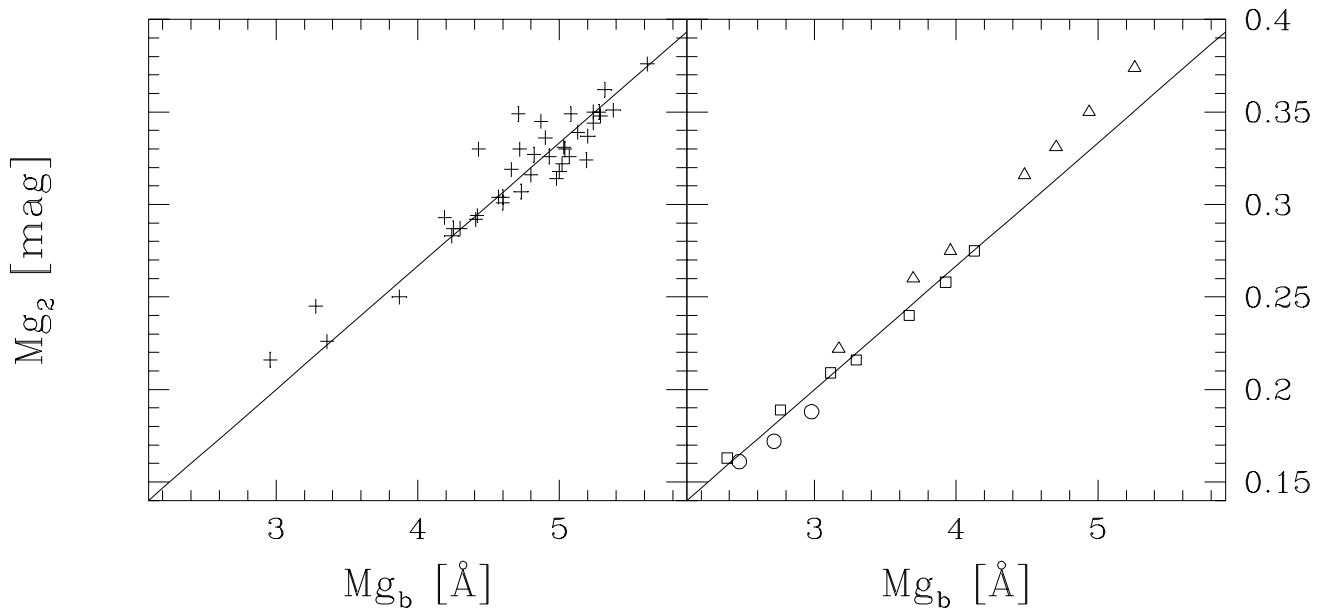,width=175mm}
\caption{Mg$_2$ vs. \mg: left panel: observed data (Gonz\'alez 1993), right
panel: calculated values (Worthey 1994) for different ages (1.5, 2, 3, 6, 8,
12, and 17 Gyrs) and metallicities ($\circ\!\!:\lg Z/Z_{\sun}=-0.5$, \oquad:
$\lg Z/Z_{\sun}=0$, \odrei: $\lg Z/Z_{\sun}=0.5$). The
straight line corresponds to Mg$_2 = $ \mg $/15$.}
\label{fig_mg2mgb}
\end{figure*}

All dynamically hot stellar systems show the same mean relationship between
central Mg$_2$ absorption and central velocity dispersion ($\sigma_0$) (e.g.
\pcite{DLBDFTW87}, \pcite{BBF93}). Although these systems comprise four
orders of magnitudes in mass and luminosity (ranging from the bulges of S0--
and spiral galaxies up to the giant ellipticals) and their Mg$_2$ equivalent
widths differ by up to $0.35\mdo$, the scatter about the mean \mgzsr\ is
very low. The Mg$_2$ index as defined by the Lick system
of absorption indices \cite{FFBG85} measures the absorption of the MgH
molecular band and the Mg\,{\sc i} triplet around \loap\ 5173\,\AA,
whereas the \mgi\ measures only this triplet with respect to an adjacent
pseudo-continuum. For reasons described below, the Mg$_2$ index can not be
determined with the same accuracy as \mg\ in our target galaxies at \rv s of
$z\approx0.37$. In order to compare the Mg absorption of the distant \es\
to published Mg$_2$ measurements of local \es\
the Mg$_2$ values have to be converted to \mg.
For the synthetic \mg\ and Mg$_2$ values calculated by \scite{Worth94} for
simple stellar populations ({\it SSP}) with ages between 1.5 and 17~Gyrs, we
find the following linear transformation (see Fig.~\ref{fig_mg2mgb}):
\begin{equation}
\label{gl_mgzmgb}
\mbox{Mg}_b\, /\mbox{\AA}=(14.3\ldots 15.5)\cdot\mbox{Mg}_2 /\mdo 
\end{equation}
with the smaller conversion factor for a \met\ of $\lg Z/Z_{\sun}=+0.5$, the
greater one for $\lg Z/Z_{\sun}=-0.5$. We adopted a slope of 15, a value
consistent with the observational data of a small sample of nearby \es\
\cite{Gonza93} and in agreement with the result by \scite{BFGK84}.

Throughout this paper, we will use some 60 \egs\ in the Coma and Virgo
clusters as a comparison sample. A principal components analysis of
the 7~Samurai data \cite{DLBDFTW87} of these \es\ yields as best fit to the
\mgsr:
\begin{equation}
\label{gl_mgbs}
\mbox{Mg}_b/\mbox{\AA} = 2.7 \lg (\sigma_0/(\mbox{km s}^{-1})) - 1.65
\end{equation}
The brighter \es\ ($\lg \sigma_0 \ge 2.3$) show a very low intrinsic scatter:
\begin{equation}
\label{gl_mgbdisp}
\sigma_{\rm intr} (\mbox{Mg}_b) = 0.16\, \mbox{\AA}
\end{equation}
\begin{figure}
\psfig{figure=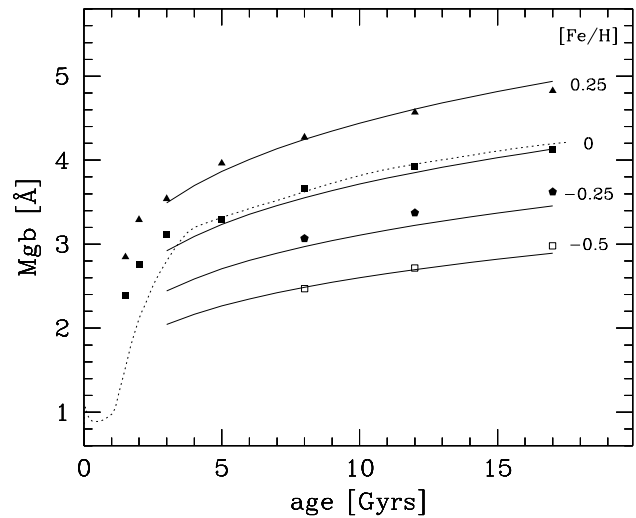,width=85mm}
\caption{Dependence of the \mgi\ on \met\ and age: symbols
represent values given by Worthey (1994), the dashed line corresponds to the
Bruzual\&Charlot (1996) model for solar metallicity, solid lines follow
equation~(\ref{gl_mgbtZ}).}
\label{fig_mgbtZ}
\end{figure}
The strength of the \mg\ absorption for a single stellar population is driven
mainly by metallicity and age. A bivariate polynomial fit to Worthey's {\it
SSP} values yields the following dependence for metallicities 
$-2<\lg Z/Z_{\sun}<+0.25$ and ages $t>3$~Gyrs, see Fig.~\ref{fig_mgbtZ}:
\begin{equation}
\label{gl_mgbtZ}
\lg \mbox{Mg}_b = 0.20 \lg t + 0.31 \lg Z/Z_{\sun} + 0.37
\end{equation}
For solar metallicity and ages $t>12$~Gyrs, $\partial \lg
\mbox{Mg}_b/\partial \lg t$ might be as low as 0.15. The same slopes are
derived for the \scite{BC97} models showing that the proportionality factors
in equation~(\ref{gl_mgbtZ}) are robust and do not depend on the \pop\ models 
(see also \pcite{Bruzu96}). Only the zeropoint is more uncertain but this
causes no problem at all because only relative changes will be considered in
the following.

The tight correlation between \mg\ and \veldis\ $\sigma_0$ of local \es\
constrains both the relative scatter in mean age ($\Delta t/t$) and the
relative scatter in mean \met\ ($\Delta Z/Z$). For the brighter \es\ in the
Coma cluster, e.g., equations \ref{gl_mgbs}, \ref{gl_mgbdisp} and
\ref{gl_mgbtZ} yield:
\begin{equation}
\label{gl_tZdisp}
\Delta t/t < 0.17 \quad \mbox{and} \quad \Delta Z/Z < 0.11
\end{equation}
This narrow constraint on the age spread of cluster \es\ implies that they
did not form continuously at the same rate but that there was a rather short
formation epoch of these galaxies. If, e.g., the majority of \es\ were
formed 12~Gyrs ago, then the scatter in age would be about 2~Gyrs.

Measuring absorption line strengths or colours alone in distant galaxies
would not allow to unambiguously determine their ages because stellar
population models have shown that effects of age and \met\ can compensate
each other (the so-called age--\met\ degeneracy, see e.g. \pcite{Worth94}.
But by comparing the \mgsr s at different redshifts relative mean ages of
cluster \es\ can be obtained, a method we dubbed \mgst\ \cite{BZB96}. This
is because the maximum scatter in $\Delta Z/Z$ for a given $\sigma_0$ is
constrained to less than 11 per cent (eq.~(\ref{gl_tZdisp})). If most of the
\egs\ evolve only passively between intermediate redshifts and today, i.e.
if no dissipative major merger occured during the last few Gyrs that could
have disturbed the \veldis\ or the \mg\ absorption (via a burst of star
formation), then any reduction of the \mgl\ of an intermediate redshift
elliptical compared to the mean value of the local sample at the same
\veldis\ is due only to its younger age.

Unlike those methods that use
\lum\ or surface brightness to determine the evolution of \es, the \mgst\
is independent of any k--corrections and corrections for extinction, both of
which can be a source of systematic errors. In addition, the influence of
the initial mass function on the amount of evolution derived from the \mgst\
is negligible, because \mg\ is determined mainly by the temperature of the
turn-off stars and not by the total number of giant stars. A further
advantage of the \mgst\ is the ability to control the selection of the
different galaxy samples. Knowing the \veldis\ of an elliptical galaxy
it is possible to estimate its mass and, therefore, to study whether there
is any significant difference in the mass distributions of the samples. Such
a difference would distort the results, because a sample with a higher
number of very massive galaxies, e.g., would also have a higher mean \met.

\section{Sample Selection}

The `passive evolution' model of stellar \pop\ predicts that observable
characteristics of \es\ like luminosity, colour and line indices change
slower and slower with time after about 3~Gyrs. Therefore, significant
differences in these parameters compared to today's \es\ are expected only
at intermediate redshifts ($z>0.3$). But, at $z=0.3$, e.g., the \mg\ triplet
is redshifted to $\lambda = 6725$\,\AA\ and falls already into that
wavelength range where any spectrum is dominated by numerous and strong
tellurial emission lines. In that range, the continuum of a typical giant
elliptical is on average ten times lower than the mean flux level of the
night sky, see upper panel of Fig.~\ref{fig_himgalf}. The situation is
even worsened by the existence of many variable absorption bands of water
vapour. Thus, maximum \snr\ of the \mga\ can be achieved only for small
redshift bins where the influence of the earth's atmosphere is lowest. The
first of such ideal redshift bins is given for $0.358<z<0.380$, if the \hb,
\mg, \fef\ and \fes\ indices are to be determined. At these redshifts,
\spec ally classified galaxy members have been published only for two
clusters: \abell\ (\pcite{MSFM88}, \pcite{PK91}) and \cl\
\cite{DG92}. To increase the number of \egs\ suitable for our investigation,
we carried out a \phot\ campaign of a sample of 20 clusters with estimated
redshifts of $z\approx 0.37$ in the $V$, $R_c$ and $I_c$ bands. This study
will be published in detail in another paper (see also \cite{Ziegl96}).
Cluster galaxies were then classified according to their $(V-R)$ and $(V-I)$
colours. Compared to the
\spec\ classification of galaxies in \abell\ and \cl\ the success rate for
identifying E/S0 galaxies correctly was about 85 per cent. In this paper,
\spec\ data will be presented of the three clusters \abell\ ($z=0.375$),
\cl\ ($z=0.377$) and \ms\ ($z=0.372$). Out of each cluster the brightest
\es\ and a number of less luminous ones were selected for \spec\
observation.

\section{Observations}

Spectroscopic observations were done during several campaigns using the 3.5m
telescope on Calar Alto and the 3.6m telescope at ESO.

During five runs on Calar Alto, a Boller\&Chivens twin spectrograph was used
at the Cassegrain focus. The grating T04 (600 lines mm$^{-1}$, dispersion:
72~\AA\ mm$^{-1}$) of the red channel yielded equal efficiency at all
observed wavelengths. The spatial resolution of the CCD was 0.9 arcsec
pixel$^{-1}$. Using a longslit, at least two galaxies could be observed at
the same time, leaving enough space for the sky, which is essential for an
accurate sky subtraction. The redshifted \es\ were observed in the
wavelength range $\lambda\lambda=6400-8000$\,\AA\ with a slit width of 3.6
arcsec, chosen to collect as much light as possible and to minimize
positioning problems. Comparison stars were observed at the corresponding
rest frame wavelengths $\lambda\lambda=4500-6100$\,\AA\ with a slit width of
2.4 arcsec, so that the spectra of both the galaxies and the stars had the
same instrumental broadening of ca. 100\,km s$^{-1}$. This arrangement is well
suited for the determination of \veldis s of \egs\ having 
$\sigma\approx 200 \ldots 300$\,km s$^{-1}$.

During two nights at the ESO 3.6m telescope, multi-object spectroscopy was
achieved with the EFOSC1 focal reducer using the grism with the lowest
available dispersion ({\sc Red150}, 120~\AA\ mm$^{-1}$). The spectra had a
lower \snr\ than those obtained at Calar Alto, mainly because the slitlets,
produced by punching round holes in a row into the multi-object mask, had a
stamp-like boundary structure which severely affected the sky subtraction.
Together with the rather high instrumental broadening of ca. 190\,km
s$^{-1}$ (using the smallest available punch head) this resulted in the data
being useful only for a comparison check with the Calar Alto data.

Because the observed galaxies have rather low apparent magnitudes
($18\mdo<R_c<20\mdo$), a total exposure time between 8 and 12 hours was
necessary to achieve at least a \snr\ $S/N>40$~\AA$^{-1}$ at the Calar Alto
3.5m telescope. These long integration times were realized by adding up
several frames with exposures of 1 to 1.5 hours. In addition, a few white
dwarf and some red giant stars \cite{FFBG85} were observed for the purpose
of flux calibration, water vapour correction, kinematical analysis and
calibration of absorption line strengths to the Lick--system.

\section{Data}

\subsection{Photometry}

For the present study the photometry had the task to yield information about
the relative exact positions, the extendedness and the energy distribution
of objects in the cluster field. For about 30 objects of each cluster the
light profile was fitted by a Gaussian to derive the center positions and
the FWHM values which enable us to discriminate between stellar and extended
objects. Intensities were measured within two concentric circles around each
object. The inner circle comprised most of the object flux whereas the outer
ring consisted mainly of sky flux. In this way, sky subtraction was achieved
accurately. The observed standard stars allowed exact flux calibration via
airmass correction and colour transformation whereas extinction correction
was applied according to
\scite{BH84}. The overall error of the magnitudes was estimated to be of
order $0.1\mdo$.

The combined data allowed the selection of candidate \egs\ for the follow-up
spectroscopy according to their $(V-R)$ and $(V-I)$ colours. The data and
contour plots of \abell, \cl\ and \ms\ are given in the appendix.

\subsubsection{Absolute Magnitudes}
\label{sec_phot}

In addition to the \mgst, the evolution of \egs\ can also be studied using
the correlation between their luminosity and \veldis\  \cite{FJ76}. To
establish the \fjr\ of the \es\ at $z=0.37$ their {\it absolute} restframe
magnitudes $(B\,V\,R)_{\rm rest}$ must be determined from their observed
{\it apparent} ones $(V\,R\,I)_{\rm obs}$. This is achieved by
transforming first the {\it aperture} magnitudes into {\it total} ones from
which then the distance modulus ($dm$) and the k--correction ($k_{\rm cor}$)
are subtracted.
\begin{equation}
\label{gl_abs}
M_i = m_{i,\rm tot} - dm - k_{i,\rm cor}, \quad \quad i=B,V,R,I
\end{equation}
\begin{figure}
\psfig{figure=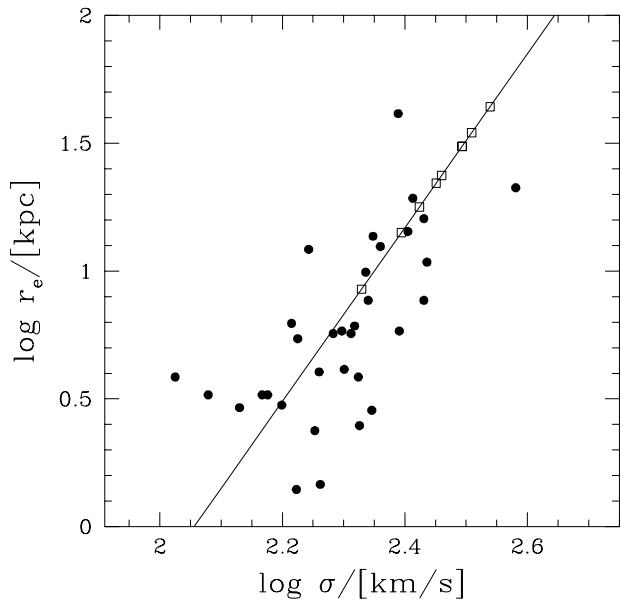,width=85mm}
\caption{Correlation between effective radius $r_e$ and \veldis\ $\sigma$ of
\egs\ ($\bullet$ measured values of Coma \es, \oquad\ calculated values for 
\abell\ according to equation~(\ref{gl_resigma}) (straight line) as derived 
from the FP relations).}
\label{fig_resigma}
\end{figure}
The ratio of aperture radius ($r_a$) to the galaxy's effective radius
($r_e$) is the important factor when extrapolating aperture magnitudes to
total ones ($r_a=\infty$). Because the $r_e$ of the faint distant galaxies
can not be measured with seeing limited groundbased photometry we estimated
these values according to the correlation between $r_e$ and $\sigma$ which can
be deduced from the {\it fundamental plane} relations:
\begin{equation}
\label{gl_resigma}
\lg (r_e/{\rmn kpc}) =  3.4 \lg (\sigma/({\rmn km\:s^{-1}})) - 6.990
\end{equation}
This relation is a rather good approximation for \es\ having $\sigma>150$~km
s$^{-1}$, see Fig.~\ref{fig_resigma}. The application of the FP relations
to the distant \es\ is justified by the recent confirmation of a FP at
intermediate redshifts ($z\approx0.4$, \pcite{DF96}). To transform the $r_e$
to apparent diameters a specific cosmology has to be chosen. For a
$\Lambda=0$ Universe, the apparent $r_e$ in arcseconds is given by
\cite{Matti58}:
\begin{equation}
\label{gl_re_as}
\frac{r_{e,\rm as}}{\rmn arcsec} = \frac{1}{3600} \frac{\pi}{180} 
\frac{r_e}{\rmn Mpc}
\frac{H_0 \,q_0{}^2 (1+z)^2 /c}{q_0 z + (q_0-1) (\sqrt{1+2q_0z} -1)}
\end{equation}
The aperture correction ($a_{\rm cor}$) can be calculated with a growth
curve ($f$) based on the $r^{1/4}$ law \cite{Vauco62} of the mean projected
light profile of \egs:
\begin{equation}
\label{gl_apcor}
a_{\rmn cor}(x(z,q_0,H_0)) /\mdo = -2.5 \lg f(x)
\end{equation}
%
%
where $x = \frac{r_a}{r_{e,\rm as}}, y = 7.668\,x^{1/4}, f = 1-b\,e^{-y}$
and $b = 1+\sum_{n=1}^7 \frac{y^n}{n!}$.
The influence of $H_0$ and $q_0$ on $a_{\rm cor}$ is increasing with $r_e$.
For $H_0=50$~km~s$^{-1}$~Mpc$^{-1}$ and $q_0=0.5$ the aperture corrections
for the observed galaxies lie between $0.2\mdo$ ($r_e=10$~kpc) and $1.1\mdo$
($r_e=40$~kpc). A typical error in $a_{\rm cor}$ of $0.2\mdo$ was estimated
from the scatter of the local \rsr\ by changing $r_e$ by a factor of 3.
Therefore, the uncertainty in the estimate of the effective radii determines
the accurateness of the aperture corrections. The values of the total
(apparent) magnitudes ($m_{\rm tot} = m_{\rm obs} - a_{\rm cor}$) change
little with respect to this uncertainty when the aperture size ($r_a$),
within which the intensities are measured, is varied. For this reason, the
error in the determination of the sky background due to a wrong choice of
apertures is smaller than the error introduced by the estimate of $r_e$.

The next step in the determination of absolute magnitudes is the
calculation of the luminosity distance ($d_L$) of the galaxies at $z=0.37$.
This is done similarly to equation~(\ref{gl_re_as}):
\begin{equation}
\label{gl_dl}
\frac{d_L}{\rmn Mpc} = \frac{c}{H_0 \,q_0{}^2} \,\left(q_0 z + (q_0-1) 
(\sqrt{1+2 q_0 z} -1) \right)
\end{equation}
Then, the distance modulus is just:
\begin{equation}
dm/\mdo = 5 \lg (d_L(z,q_0,H_0)) + 25
\end{equation}

To determine the k--correction we have created model spectra using
\pop\ \cite{BC97} that matched luminosities and colours of the observed \es\
of \abell. Model galaxies contained a
stellar population which was formed within a 1~Gyr burst and evolved only
passively thereafter.
The model spectra allowed the measurement of both apparent
`observed' magnitudes and absolute restframe magnitudes yielding the
k--corrections according to equation~(\ref{gl_abs}). For a \rv\ of $z=0.37$ we
found the following average values with uncertainties of $0.05\mdo$:
\begin{eqnarray}
k_{\rm cor} (B_{\rm rest},B_{\rm obs}) & = & -1.78 \mdo
\label{gl_bb} \\
k_{\rm cor} (B_{\rm rest},V_{\rm obs}) & = & -0.22 \mdo
\label{gl_bv} \\
k_{\rm cor} (V_{\rm rest},V_{\rm obs}) & = & -1.15 \mdo
\label{gl_vv} \\
k_{\rm cor} (R_{\rm rest},R_{\rm obs}) & = & -0.58 \mdo
\label{gl_rr} 
\end{eqnarray}
To study the influence of the initial mass function (IMF) on the evolution
of the absolute magnitudes we created model galaxies having different
$x$--values of the standard Salpeter parametrisation 
($\Phi(m)\propto m^{-(x+1)}$).
For the $B$~magnitude, e.g., we found the following dependence, which is a
good approximation for ages $t>1.5$~Gyrs:
\begin{equation}
\label{gl_bageimf}
\Delta B/\mdo \approx 3.35 \,\Delta \lg (t/{\rmn yr}) \,\cdot\, 
[1-0.24\,(x-1.35)]
\end{equation}
This formula is in good agreement with the one given by \scite{Tinsl80}. For
a flat Universe ($\Omega=1$) the time dependence can be transformed into a
function of \rv, because the scale factor ($R=1/(1+z)$) then depends on time
as $R\propto t^{2/3}$:
\begin{equation}
\label{gl_bzimf}
\Delta B/\mdo \approx 2.18 \ln (1+z) \,\cdot\,  [1-0.24\,(x-1.35)]
\end{equation}
According to this formula a deviation of 1 from the Salpeter value of $x$
(1.35) results in a change of $0.16\mdo$ in $B$ at the redshift of the
observed galaxies ($z=0.37$).

\subsection{Spectroscopy}
\label{sec_spec}

All spectra were carefully reduced using standard techniques (see, e.g.,
\pcite{BSG94}): bias and dark subtraction, flatfield division, cosmics
removal, sky subtraction, logarithmic wavelength calibration, extraction of
a one-dimensional spectrum and, finally, summation of the individual spectra
per galaxy.

\begin{figure*}
\psfig{figure=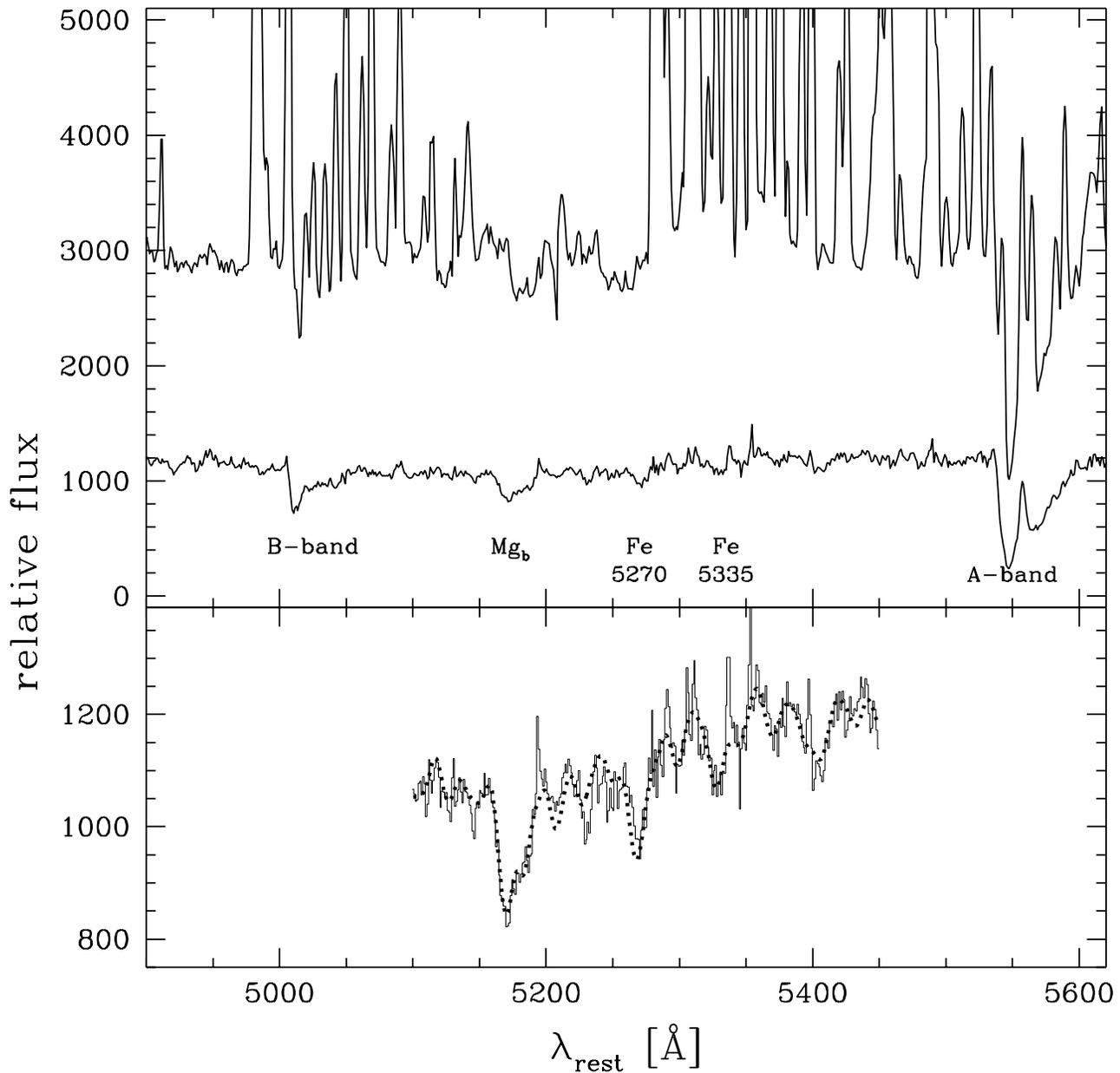,width=175mm}
\caption{Upper panel: rest--frame spectra of an elliptical
galaxy in the cluster \abell\ before and after sky subtraction. The Mg
absorption around $\lambda_0=5170$\,\AA\ can be readily seen in the lower sky
subtracted spectrum. (The features at $\lambda_0\sim5000$ and
$\lambda_0\sim5550$\,\AA\ are the blueshifted atmospheric B- and A-band,
resp.). Lower panel: fit (dotted line) to the galaxy spectrum using a
superposition of spectra of nearby \es.}
\label{fig_himgalf}
\end{figure*}

Because the galaxy spectra are heavily dominated by sky emission lines at
the observed wavelengths (see Fig.~\ref{fig_himgalf}), the following
procedure was applied for the sky subtraction: In two windows (each 10 arcsec
wide) neighbouring the galaxy spectrum the intensity distribution of each
column of the CCD frame (representing the spatial dimension) was fitted by a
polynomial. By interpolating the fit functions over the rows of the galaxy
spectrum a model image of the sky was created which then was subtracted from
the raw image. In this way, the flux level of the background was reduced to
a few percent of the galaxy's continuum level except for those regions that
contained originally very strong emission lines. In
Fig.~\ref{fig_himgalf} it can be seen that the sky level is about three
times higher than the galaxy's flux at the \mga\ whereas this ratio is about
ten at the \fes\ line. Nevertheless, even this line is well approximated by
a fit using the superposition of five nearby \egs\ of different line
strengths (see below).

Extraction of the one-dimensional spectra was performed by an algorithm
described by \scite{Horne86}. Each row of the galaxy spectrum is weighted
and added up in a way to achieve maximum \snr\ for each pixel in the
resulting one-dimensional spectrum. At the same time, cosmics are removed
by analysing the profile perpendicular to the dispersion axis. Those pixel
values that exceed the median of neighbouring pixels by more than a given
threshold will be replaced by this median value. To determine the average
\snr\ of the whole spectrum its power spectrum was compared to synthesized
noisy power spectra of a broadened comparison star. The observed galaxies
have an $S/N$ of $30 \ldots 50$ per \AA ngstrom after sky subtraction.

Kinematic parameters (radial velocities and \veldis s) of the distant \egs\
were determined by two different methods. The first one was based on the
Fourier correlation quotient (FCQ) analysis \cite{Bende90a}. To check the
reliability of the determined values the FCQ analysis was repeated using
five different template stars of different spectral types and applied to
different wavelength regions. The other method was a direct fitting
procedure and was also applied to various parts of the galaxy spectrum.
Here, essentially, several broadened spectra of either local \es\ or stars
were superposed on each other in a way to give an optimal fit to the
spectrum. By systematically varying the broadening factor for the input
spectra the \veldis\ and its error could be estimated (for a detailed
description see \cite{Ziegl96}). All procedures were
inspected visually and the results assigned a quality mark. In this way, the
kinematic parameters were determined to an accuracy of about 10 per cent on
average.

Line strengths of \hb\ (\lo\ 4861\,\AA), \mg\ (\lo\ 5173\,\AA), Mg$_2$
(\loap\ 5175\,\AA ; i.e. MgH+\mg), \fef\ (\lo\ 5269\,\AA) and \fes\ (\lo\
5328\,\AA) were measured according to the Lick system \cite{FFBG85}. But
only the \mga\ could be determined accurately; measurements of the other
\lst s are far less accurate, because these lines are affected by problems
of sky subtraction and/or emission lines. With respect to \hb\ there is no
means to correct for a possible contamination of the absorption by emission,
because for most of the galaxies the emission line of [O\,{\sc iii}] (\lo\
5007), which is usually used for this correction, is redshifted to the
B--band, a very strong telluric absorption band. The low \hb\ index of
galaxy A20 of \abell\ (see appendix for the nomenclature) that shows
[O\,{\sc iii}] in emission demonstrates the possibility that \hb\ might be
partially filled by emission in the other galaxies. The \abs s of \fef\ and
\fes\ are in most cases rather noisy because they lie in that region of the
spectrum which is dominated by very strong sky lines (see
Fig.~\ref{fig_himgalf}). Here, the residuals after sky subtraction are so
large that they prevent any reliable measurement. This situation is worst
for \fes. In addition, there are several weak water bands so variable that
they can not be corrected for with a spectrophotometric standard star. The
same problem arises for Mg$_2$, because its red continuum window coincides
with the \fes\ line. In order to compare the linestrengths of the distant
galaxies with the Lick system the effects of different spectral resolution,
broadening by the velocity dispersion, and redshift had been taken into
account. Absorption line strengths of galaxies with high \veldis\ are
systematically underestimated in the Lick system due to the fixed continuum
windows. But this effect can easily be corrected by simulations with
broadened template stars. The \lst s of \mg\ and \hb\ of all observed
galaxies are tabulated in the appendix (Table~\ref{tab_spek}). The \mgl s
could be determined to an accuracy of about 5 per cent on average.

\begin{figure*}
\psfig{figure=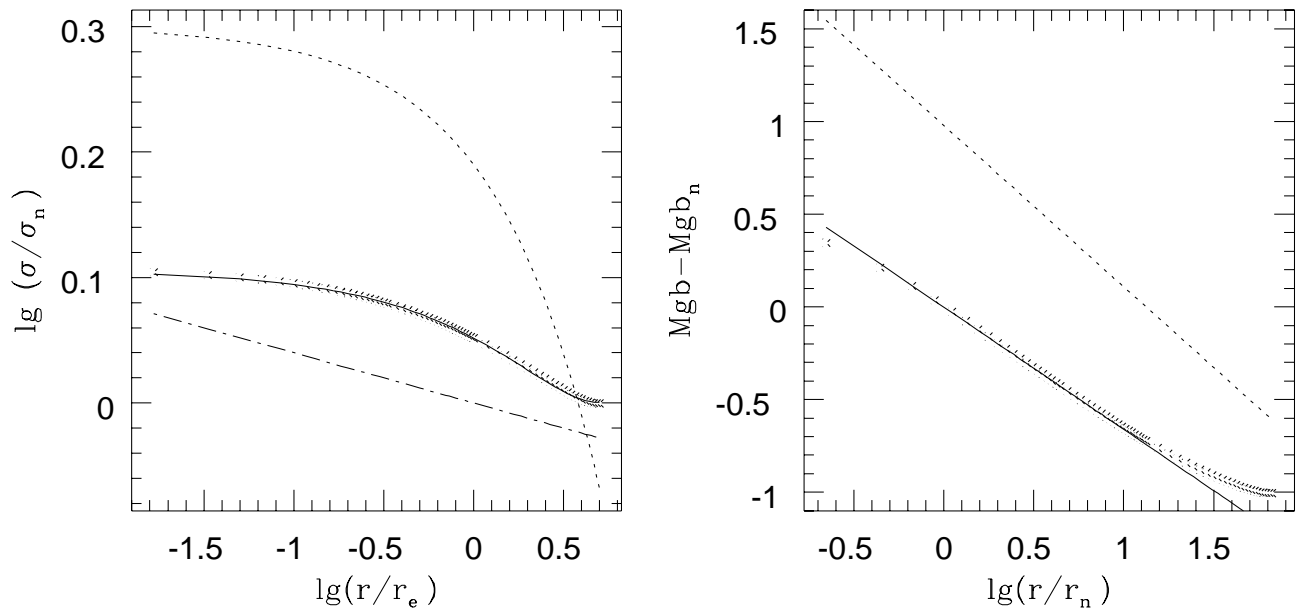,width=175mm}
\caption{Radial profiles of the mean values for
$\langle\lg\sigma\rangle$ (left panel) and $\langle\mbox{Mg}_b\rangle$ 
(right panel)
(see eq.~(\ref{gl_meanval})): crosses = mean values as computed by the
simulation; solid line = fit to these values; dotted line = radial gradient
for $\lg\sigma$ (eq.~(\ref{gl_siggrad})) and \mg\ (eq.~(\ref{gl_mgbgrad}));
dot-dashed line = logarithmic gradient for $\lg\sigma$ chosen by other
authors.}
\label{fig_grads}
\end{figure*}

The distant galaxies are so faint that all the observed light must be
combined in a one-dimensional spectrum leading to mean values of the
extracted parameters weighted by luminosity. Because \egs\ have radial
gradients in both \veldis\ and \lst s, the effect of different aperture size
must be taken into account when comparing distant to local galaxies. A much
larger part of the galaxy will be averaged for the distant \es\ than for the
nearby ones. Thus, the quasi {\it integral} values of the $z=0.37$ \es\ must be
transformed into quasi {\it central} values as they have been observed for our
comparison sample of Coma and Virgo \es. To study the dependence of \mg\ and
$\sigma$ on aperture size we made simulations with a model galaxy whose
surface brightness followed the de Vaucouleurs law and assuming \mg\ and
$\sigma$ to be constant along isophotes. The mean values were calculated
according to the following formula:
\begin{equation}
\label{gl_meanval}
\langle X \rangle \quad = \int\limits_{aperture} \!\!\!\!\! I(r) X(r) \, dA 
\left/ \!\! \int\limits_{aperture} \!\!\!\!\! I(r) \, dA \right.
\end{equation}
with $I(r)$ and $X(r)$ being the radial profiles of the intensity and \mg\
or $\log \sigma$, respectively and $A$ the total area of the aperture.
Ideally, the functions Mg$_b(r)$ and $\log \sigma(r)$ should be determined
from data of several \es\ with $r$ sampled out to at least five effective
radii ($r_e$). In the case of \mg\ we deduced the following profile based on
a study of 114 \egs\ out to about $3\,r_e$ \cite{GG95} (using
eq.~(\ref{gl_mgzmgb}) to transform Mg$_2$ into \mg):
\begin{equation}
\label{gl_mgbgrad}
{\rm Mg}_b = -0.87 \lg (r/r_e) + c
\end{equation}
From a yet unpublished investigation of nearby \es\ by Saglia et al. with
data out to $r\approx2.5\,r_e$ we find the following profile for $\log\sigma$:
\begin{equation}
\label{gl_siggrad}
\lg \sigma (r) = -0.11 \, (r/r_e)^{3/4} + c
\end{equation}
This profile falls off considerably steeper for large $r_e$ than previously
published profiles based on data with a smaller radial extent (e.g.
\pcite{JFK95}). The power law function of $\log \sigma(r)$ leads to a
dependence of the aperture correction on the galaxy's effective radius
whereas the correction is independent of $r_e$ in the case of the
logarithmic function of Mg$_b(r)$. Both the input functions $X(r)$ and the
determined profiles of the mean values $\langle X(\le\!r)\rangle(r)$ are
illustrated in Fig.~\ref{fig_grads}. Because we can not determine the
effective radii of all our distant galaxies accurately we chose a value of
30 arcsec as a first approximation in the present study (see
\pcite{Ziegl97b} for the subsample with measured $r_e$ from {\it HST}
images). To transform our measured data of the
distant \es\ to the apparent diameters of Coma and Virgo \es\ and the
aperture size used for their observations \cite{DLBDFTW87} we applied a mean
aperture correction of $\Delta \lg \sigma = 0.042$ and $\Delta {\rm
Mg}_b=0.60$.

Using the same simulations we studied the influence on the aperture
correction caused by the ellipticity of the galaxy (E1 -- E7), the deviation
of position angle from slit angle (0 -- 90 degrees), the offset between
galaxy center and slit position (0 -- 2 arcsec) and the ratio of the sides
of a rectangular slit (1 -- $\infty$). It turned out that the variations of
the aperture corrections amounted in the most cases only to $1 \ldots 2$ per
cent and could be therefore neglected. Only $\Delta \lg \sigma$ varied up to
10 per cent in extreme cases of high eccentricity or large misplacement of
the slit.

\section{Results}

\subsection{The \bmgs\ relation at $\bmath{z\approx0.37}$ and the ages of
\egs} 

\begin{figure*}
\psfig{figure=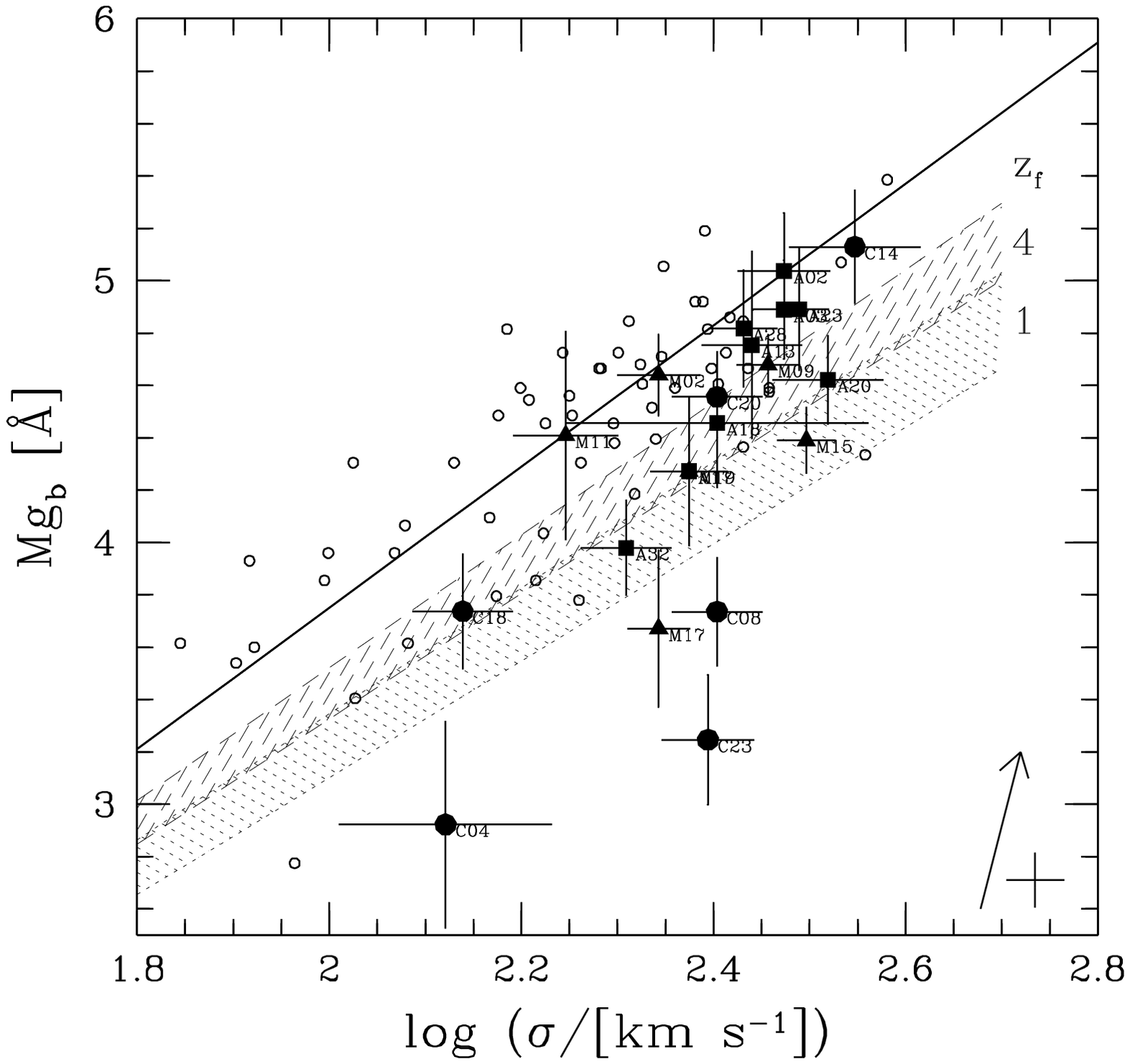,width=175mm}
\caption{ Mg$_b-\sigma$ pairs at $z=0.37$ (big symbols with
errorbars, labels see appendix) compared to the local \mgsr\
(solid line: eq.~(\ref{gl_mgbs}), small circles: Coma and Virgo \es, typical
errorbar in lower right corner). Arrow: aperture correction applied.
Hatched areas: expected \mgs\ at $z_{\rmn obs}=0.37$ for $z_f=1$ and 4,
respectively, and different cosmologies and stellar population models.}
\label{fig_mgbs}
\end{figure*}

Here, we present data of 21 \egs\ in three clusters at nearly the same \rv:
\ms\ ($z=0.372$), \abell\ ($z=0.375$) and \cl\ ($z=0.377$). In
Fig.~\ref{fig_mgbs}, the distribution of the age and \met\ dependent \mgi\
and the internal \veldis\ ($\sigma$) of these galaxies as well as of our
comparison sample of nearby \es\ in the Coma and Virgo cluster is given. The
distant \es\ show a similar correlation between the two parameters like the
local ones, but \mg\ is lower than the mean value of the comparison galaxies
for any given $\sigma$. This can not be an artifact of our selection. The
\es\ of our sample in \abell\ have a colour cut-off $(B-V)_{\rm obs}>1.4\mdo$.
Applying k--corrections
(eqs.~\ref{gl_bb} and \ref{gl_vv}) the rest--frame cut-off is 
$(B-V)_{\rm rest} > 0.8\mdo$. This colour criterion translates into a selection 
of our galaxies with respect to \mg\ and $\sigma$, when the tight correlations
between $(B-V)$ colour and these parameters \cite{BBF93} together with 
equation~(\ref{gl_mgzmgb}) are considered:
\begin{equation}
\label{gl_bvmgz}
(B-V)\,_0 = 1.12 \,\mbox{Mg}_2  + 0.615 
\end{equation}
\begin{displaymath}
\Rightarrow \quad \mbox{Mg}_{b,\rm obs} \ge 2.5 \,\mbox{\AA} 
\end{displaymath}
\begin{equation}
\label{gl_bvs}
(B-V)\,_0 = 0.224 \lg \sigma_0  + 0.429 
\end{equation}
\begin{displaymath}
\Rightarrow \quad \lg (\sigma_{\rm obs} / ({\rmn km\:s^{-1}})) \ge 1.65
\end{displaymath}
So, we should in principle be able to detect objects with \mg\ as weak as
2.5\,\AA. The absence of objects with high $\sigma$ and low \mg\ is
therefore significant and underlines the existence of an \mgsr\ at $z=0.37$.

The reduction of \mg\ with respect to the mean local relationship is weak
but significant. A student's T--test gives a significantly lower mean value
for the distant sample than for the comparison sample. The mean reduction
$\dmgb=-0.37$\,\AA\ with an error of the mean of $0.08$\,\AA\ between
$z=0.37$ (corresponding to a look--back time $t_{\rmn lb}\approx5$~Gyrs) and
$z=0$ is so low that it can be only understood, if the distant stellar
populations are already old themselves. It is fully consistent with a pure
passively-evolutionary behaviour of
\egs\ since $z=0.37$. Significant star bursts that would change the overall
\met\ at a detectable level are ruled out because of the small evolution.
Thus, equation~(\ref{gl_mgbtZ}) can be transformed to:
\begin{equation}
\label{gl_mgbt}
\frac{\mbox{Mg}_b (z\!=\!0)}{\mbox{Mg}_b (z)} = 
\left( \frac{\mbox{age}\,(z\!=\!0)}{\mbox{age}\,(z)} \right)^{0.15\ldots 0.20}
\end{equation}
With the average values Mg$_b(z\!=\!0)=4.8$\,\AA\ and
$\dmgb(z\!=\!0.37)=-0.37$\,\AA, the distant galaxies have already
$\approx2/3$ the age of local \es. Equation~(\ref{gl_mgbt}) in connection
with the local \mgsr\ (eq.~(\ref{gl_mgbs})) can be used to derive expected
functions of Mg$_b(\sigma,z\!=\!0.37)$ for different cosmologies ($H_0,
\Omega_0$) and
\rv s of formation ($z_f$). In Fig.~\ref{fig_mgbs}, the expected locations
of \mgs\ at $z_{\rmn obs}=0.37$ are shown as hatched areas for $z_f=1$ and
4, respectively, and any combination of $\Lambda=0, q_0=0\ldots 1,
H_0=50\ldots 100 {\rmn \:km\:s^{-1}\:Mpc^{-1}}$ and $\partial \lg
\mbox{Mg}_b/\partial \lg t=0.15\ldots 0.20$.
Comparing this with the observed values it follows that
the majority of the stars of most of the \egs\ in clusters must have been
formed at \rv s $z>2$, of the most luminous galaxies probably even at $z>4$.
The estimated age is probably a lower
limit to the real formation era, because the applied aperture correction
(arrow in fig.~\ref{fig_mgbs}) followed rather conservative assumptions on
the gradients of \mg\ and $\sigma$ (i.e. too shallow gradients).

\begin{figure*}
\psfig{figure=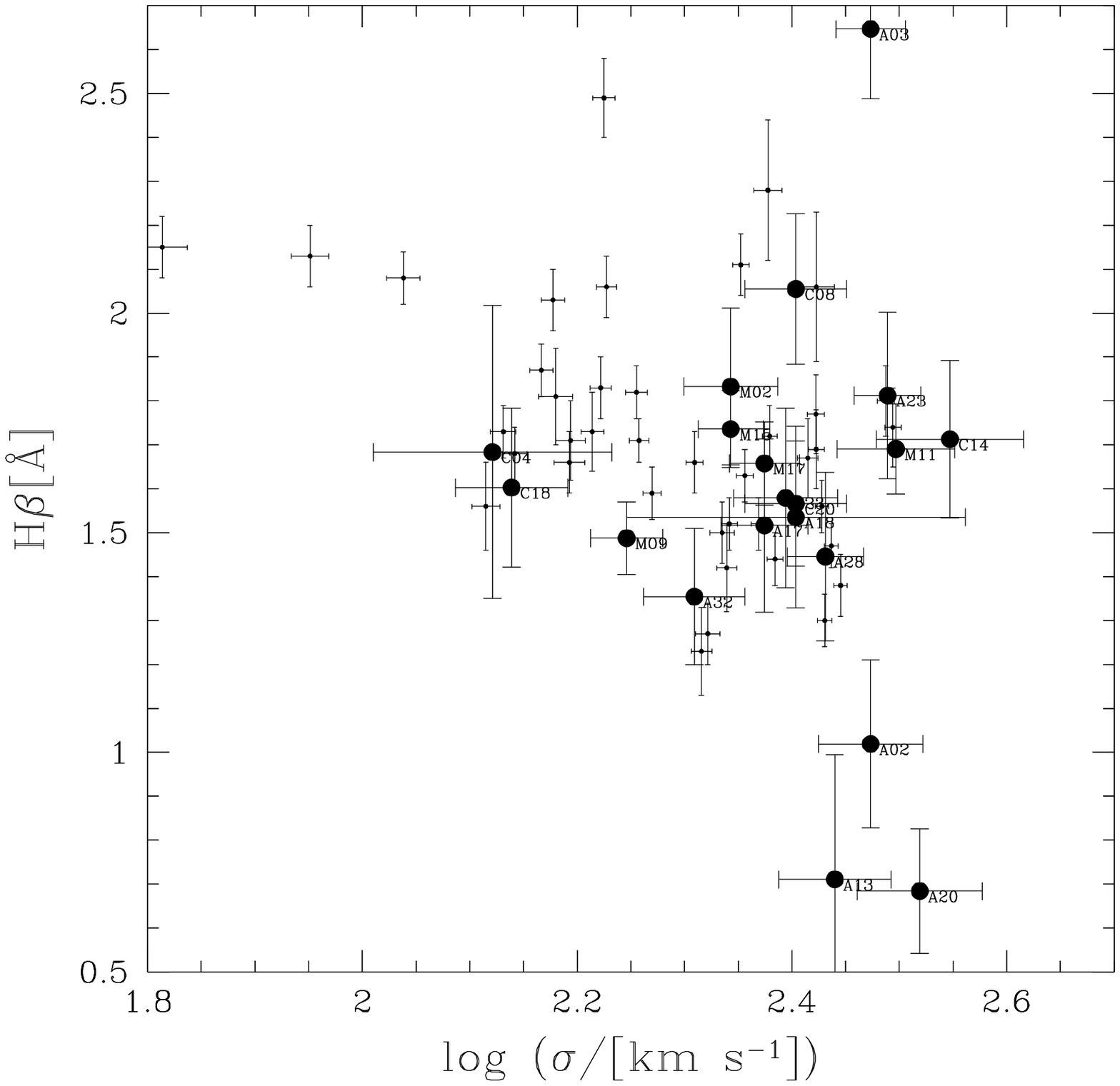,width=175mm}
\caption{ H$_\beta-\sigma$ pairs at $z=0.37$ (big symbols, labels
see table~1) compared to nearby field \es\ (small symbols, Gonz\'alez 1993).}
\label{fig_hbs}
\end{figure*}

Given the large errorbars of our current data and the very low number of
observed galaxies with low $\sigma$, it is rather speculative to comment on
a possible change of the slope of the \mgsr. Taking the data at face value
it seems that the less massive \es\ are younger than the more massive ones.
This conclusion can not be circumvented by claiming that those galaxies with
very low \mg\ for their $\sigma$ are E$+$A galaxies that had one late
starburst. Because of our colour selection, an E$+$A would enter our sample
only 2~Gyrs after its starburst when its $(B-V)$ colour returned to almost
normal. At that time, the \mgi\ has also almost reached the value it had
before the starburst. In fact, galaxy {\it A28} of \abell\ was put close to 
the E$+$A class on the basis of its high \hd\ absorption \cite{HL87}, but both
its \mg\ and \hb\ are like in normal \es\ (see Figs.~\ref{fig_mgbs} and
\ref{fig_hbs}). 

It was claimed recently on the basis of observed higher values of \hb, that
less luminous \es\ could have younger mean ages than giants \cite{FTGW95}.
Our distant galaxies do not show any correlation with the \hb\ index in the
sense that the galaxies with very low \mg\ would have very high \hb. But
remember the problems in determining \hb\ of the distant \es\ as stated in
Section~\ref{sec_spec}. Also, we applied no aperture corrections to \hb, for
nearby \es\ are found to have nearly radially constant \hb\ \cite{Gonza93}.
Fig.~\ref{fig_hbs} compares the \hb\ values of the distant \es\ to the nearby
sample. There is no significant difference between the two distributions and
the outliers can be understood in terms of peculiarities of the spectra.

\subsection{The \fjr\ at $\bmath{z\approx0.37}$ and the luminosity evolution
of \egs} 

\begin{figure*}
\psfig{figure=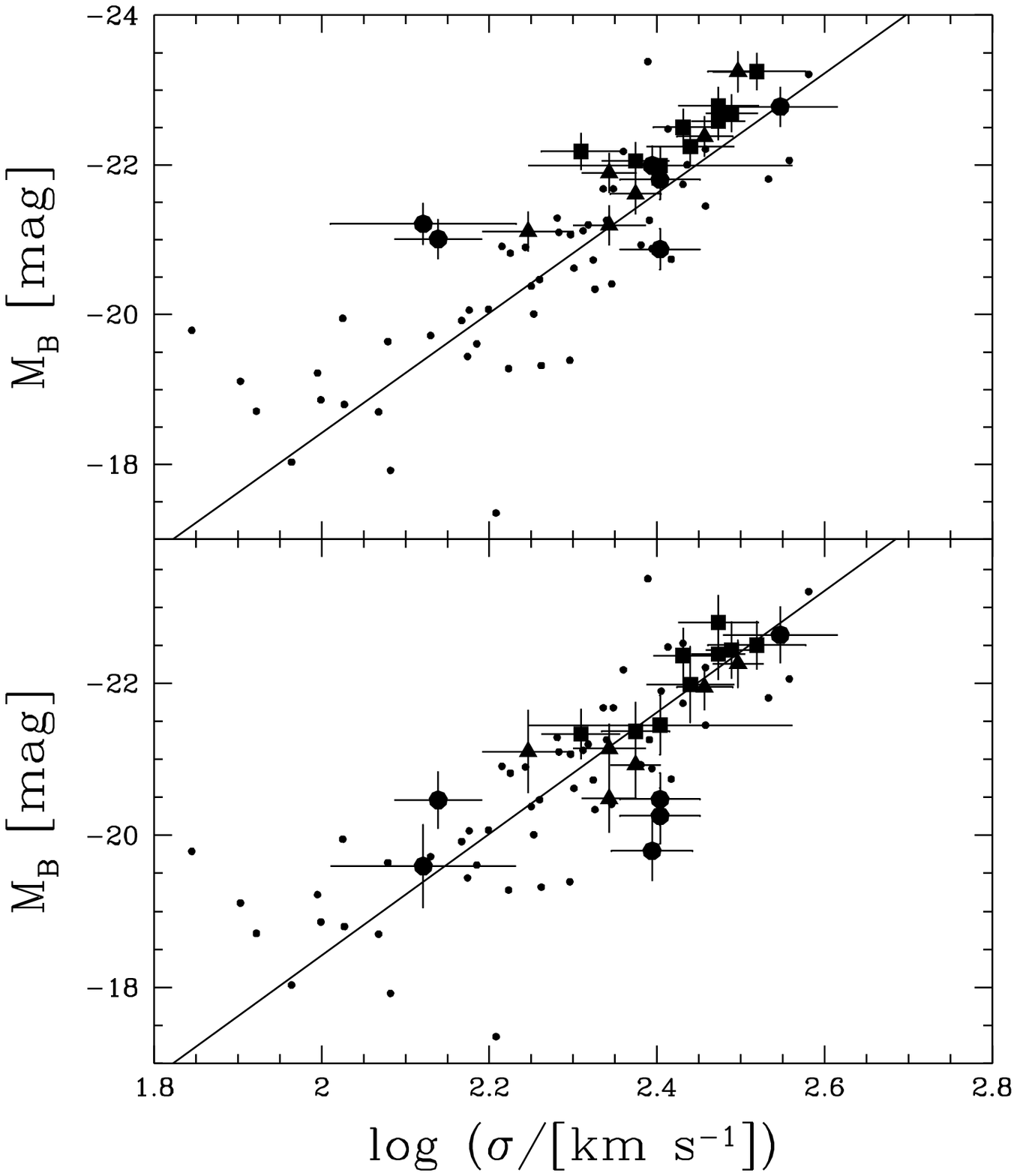,width=175mm}
\caption{ M$_B-\sigma$ pairs at $z=0.37$ (big symbols with errorbars)
compared to the local \fjr\ (solid line: (eq.~(\ref{gl_mbsfit})), small
circles: Coma and Virgo \es). Upper panel: uncorrected absolute
$B$~magnitudes ($H_0=50{\rmn \:km\:s^{-1}\:Mpc^{-1}},q_0=0.5, \Lambda=0$),
lower panel: magnitudes of the distant \es\ corrected for evolution as found
by the \mgst.}
\label{fig_fjr}
\end{figure*}

Stellar population models predict an evolution with age not only for
absorption lines but also for the luminosity. For passively evolving simple
stellar populations the increase in brightness with \rv\ is most prominent
in the $B$--band. Luminosity differences of \egs\ can be well studied on the
basis of the tight correlation between absolute blue magnitude (M$_B$) and
\veldis\ \cite{FJ76}. The very small evolution found with the \mgst\
excludes dissipative mergers that could substantially change $\sigma$
between $z=0.4$ and today. Non-dissipative mergers, that are not strictly
ruled out in passive evolution models, do also not lead to any substantial
change of the \veldis\ (e.g. \pcite{AF80}, \pcite{HHS96}). Thus, M$_B$ of
our distant galaxies can be directly compared to M$_B$ of the comparison
\es\ at the same \veldis. The determination of M$_B$ as described in
Section~\ref{sec_phot} depends first on the observational data (apparent
magnitudes, aperture correction), second on the k--correction and third on
the chosen cosmology and IMF. In the upper panel of Fig.~\ref{fig_fjr} the
data of the \es\ at $z=0.37$ (choosing $H_0=50{\rmn \:km\:s^{-1}\:Mpc^{-1}},
q_0=0.5, \Lambda=0$) are compared to data of Coma and Virgo \es\
\cite{DLBDFTW87}. A principal components analysis of the Coma data yields as
best fit:
\begin{equation}
\label{gl_mbsfit}
M_B /\mdo =  -2.42 - 8 \lg (\sigma / ({\rmn km\:s^{-1}}))
\end{equation}
The distant \es\ are on average significantly more luminous than the nearby
ones as proven by a student's T--test. The mean brightening in this example
amounts to $\dmb(z\!=\!0.37)=-0.63\pm0.10\mdo$. Now, the question arises,
whether this evolution of the luminosity is compatible with the results of
the \mgst. From stellar \pop\ models \cite{Worth94} for SSP's
(Salpeter--IMF) we find a linear relationship between $\Delta \mbox{M}_B$
and $\Delta \mbox{Mg}_b$ which is well suited for ages greater than 1.5~Gyrs
and metallicities between half and twice solar:
\begin{equation}
\label{gl_mgbmb}
\Delta \mbox{M}_B /\mdo \approx  (1.4 \pm 0.1) \cdot
\,\Delta \mbox{Mg}_b /\mbox{\AA}
\end{equation}
With this formula, the mean reduction of
$\dmgb(z\!=\!0.37)=-0.37\pm0.08$\,\AA\ translates into
$\dmb(z\!=\!0.37)=-0.50\pm0.11\mdo$. Thus, the amount of evolution of \egs\
between $z=0$ and $z=0.37$ found with the \mgst\ is in agreement with the
brightening of the galaxies as derived from the \fjr\ for the chosen
cosmology. Varying $q_0$ from 0.5 to 1 or 0 would change the absolute
$B$~magnitude of the observed \es\ by $\pm0.11\mdo$ on average (remember
that not only the luminosity distance is affected but also the effective radii
and therefore the aperture correction). A change of the slope of the IMF by
$\Delta x=\pm1$ from the Salpeter value of $x=1.35$ would result in a
shift of the $B$~magnitudes of $\pm0.16\mdo$ (eq.~(\ref{gl_bzimf})). Thus, the
current data with their errors do not indicate an unusual IMF slope and are
compatible with $q_0=0.5\pm0.5$.

The above procedure can also be turned around. Then, for each galaxy the
$B$~magnitude is corrected for evolution according to its individual
reduction of \mg\ with respect to the mean local \mgsr\ using
equation~(\ref{gl_mgbmb}). The result of this individual correction for the
luminosity evolution is shown in the lower panel of Fig.~\ref{fig_fjr}.
Now, the distributions of the distant and local \es\ are almost identical and
even the slopes are very similar. This time, the average brightening amounts
to $\dmb(z\!=\!0.37)=-0.50\pm0.14\mdo$.

\section{Conclusions}

Comparing the \mga\ index of a sample of \egs\ in three clusters at a \rv\
$z=0.37$ with the local \mgsr\ we find an average reduction
$\dmgb=-0.37\pm0.08$\,\AA. This is evidence for significant but weak
evolution of \egs\ in clusters within a look--back time of ca. 5~Gyrs. It is
compatible with the passive evolution of stellar \pop\ models
(\pcite{Worth94}, \pcite{BC97}). The mild evolution requires that the
majority of the stellar population of normal cluster \es\ was formed at
\rv s $z_f>2$. The most massive \es\ might even have a mean formation \rv\ of
$z_f>4$. This implies that star formation happened already when the Universe
was very young (for the standard cosmology $\Lambda=0,q_0=0.5,H_0=50{\rmn
\:km\:s^{-1}\:Mpc^{-1}}$, $z=4$ corresponds to $t_U=1$~Gyr only). But in the
framework of CDM-dominated hierarchical clustering giant \es\ are assembled
from smaller entities at much later times \cite{BCF97}. In order to have
nevertheless such a high mean stellar age, the merging of the protogalaxies
to a very massive elliptical galaxy must have been essentially
dissipationless without any significant new star formation. The gas content
of the merging halos must therefore have been very low in comparison with
their stellar mass. The normal mass \es\ with mean stellar ages
corresponding to $z_f=2$ ($t_U=2.5$~Gyrs) on the other side, could have
their major starforming phase at the time of their assembly. Such a
scenario corresponds to the {\it gas/stellar continuum} model
\cite{BBF93}: the gas content and, therefore, the role of dissipation of the
last {\it major merger} decreases for increasing mass of the resultant
galaxy. The epoch of formation derived here is also in agreement with
semi-analytic CDM simulations, for which the last dissipative {\it major
merger} leading to a giant elliptical occured at $z>2$ \cite{Kauff96}. If
the slope of the distant \mgsr\ is different from the local one as
marginally indicated by the present data then less luminous \es\ would be
systematically younger than the most luminous ones. This would agree with an
\hb\ analysis of a sample of nearby \es\ suggesting that the mean age of the
stellar population gets younger with decreasing luminosity \cite{FTGW95}.

We showed that the weakening of the \mgi\ of cluster \es\ at $z=0.37$
corresponds to the brightening of the $B$~luminosity by
$\dmb=-0.50\pm0.14\mdo$. This is in quantitative agreement with \pop\ of
passively evolving galaxies. Our synthesized galaxy (using \pcite{BC97}
models) that matches the observed colours of \es\ in \abell\ (see
Section~\ref{sec_phot}: k--correction) experiences an evolution of the
rest--frame $B$~magnitude of $\Delta \mbox{M}_B=-0.61\mdo$. It is also
consistent with results obtained by other groups. \scite{SCYLE96} find,
e.g., an increase of the blue luminosity by
$\Delta\mbox{M}_{AB}(B)=-0.55\pm0.12\mdo$ for early-type galaxies in the
cluster {\it MS\,1621+26} at $z=0.43$ and \scite{BSL96}
$\Delta\mbox{M}_B=-0.64\pm0.3\mdo$ in the cluster {\it CL\,0939+47} at
$z=0.41$.

The agreement of the evolutionary effects as found via the \mgsr\ and the
\fjr\ strongly supports the hypothesis that the stellar populations of \egs\
in the density environment of clusters of galaxies formed very early and in
a short period of time with no substantial new star formation between
$z=0.4$ and today. This result is not biased in the sense that we would have
chosen only that (small) fraction of the whole population of \egs\ that {\it
is} old (for a discussion see, e.g. \pcite{FD96}), because our selection
criterion did not pick up only the reddest members. The rest--frame $(B-V)$
colour cut-off of $0.8\mdo$ is well below the mean value for samples of
nearby early-type galaxies. Of course, a much bigger sample is needed to
clarify this issue in detail with selection based on \spec\ criteria and not
on colours. Another aspect to be studied in the future is the possible
dependence of the \mgsr\ on the density environment (\pcite{CD92},
\pcite{Lucey95}, \pcite{Joerg97}). The three clusters investigated
in this paper do have different richnesses but the small number of observed
galaxies does not allow us to draw statistically significant conclusions about
any dependence on density environment. The cluster with the biggest number
of observed galaxies, \abell, has a similar richness class like our
local comparison cluster, Coma.

In a follow-up paper, we combine the individual evolutionary corrections as
found via the \mgsr\ with a full fundamental plane analysis of our HST
images of the three clusters to calibrate \egs\ as standard candles for the
determination of the cosmological de-/acceleration parameter $q_0$
(\pcite{BSZBBGH97b}). Preliminary results are given in \scite{BSZ96}.

\section*{Acknowledgments}

The authors would like to thank Dr. G. Bruzual for his continous support
with his models as well as Drs. R. P. Saglia, P. Belloni, L. Greggio and U.
Hopp for many fruitful discussions. 
This work was supported by the ``Sonderforschungsbereich 375--95 f\"ur
Astro--Teilchenphysik der Deutschen Forschungsgemeinschaft'' and by DARA
grant 50\,OR\,9608\,5.





\appendix

\section{Photometric Data}

Contour plots of the clusters \abell, \cl\ and \ms\ are shown with the
observed objects marked. North is up and east is left. \abell\ was observed
with the ESO NTT using the EMMI focal reducer with a spatial resolution of
0.268~arcsec~pixel$^{-1}$ and a field of view of ca. 9\arcmin\ x 9\arcmin.
\cl\ and \ms\ were observed with the 2.2m-telescope on Calar Alto using a
CCD camera at the Cassegrain focus with a spatial resolution of
0.281~arcsec~pixel$^{-1}$ and a field of view of ca. 4.5\arcmin\ x
4.5\arcmin. Tables give magnitudes in the standard Kron--Cousins filter
system \cite{Besse83}. 

\begin{figure*}
\psfig{figure=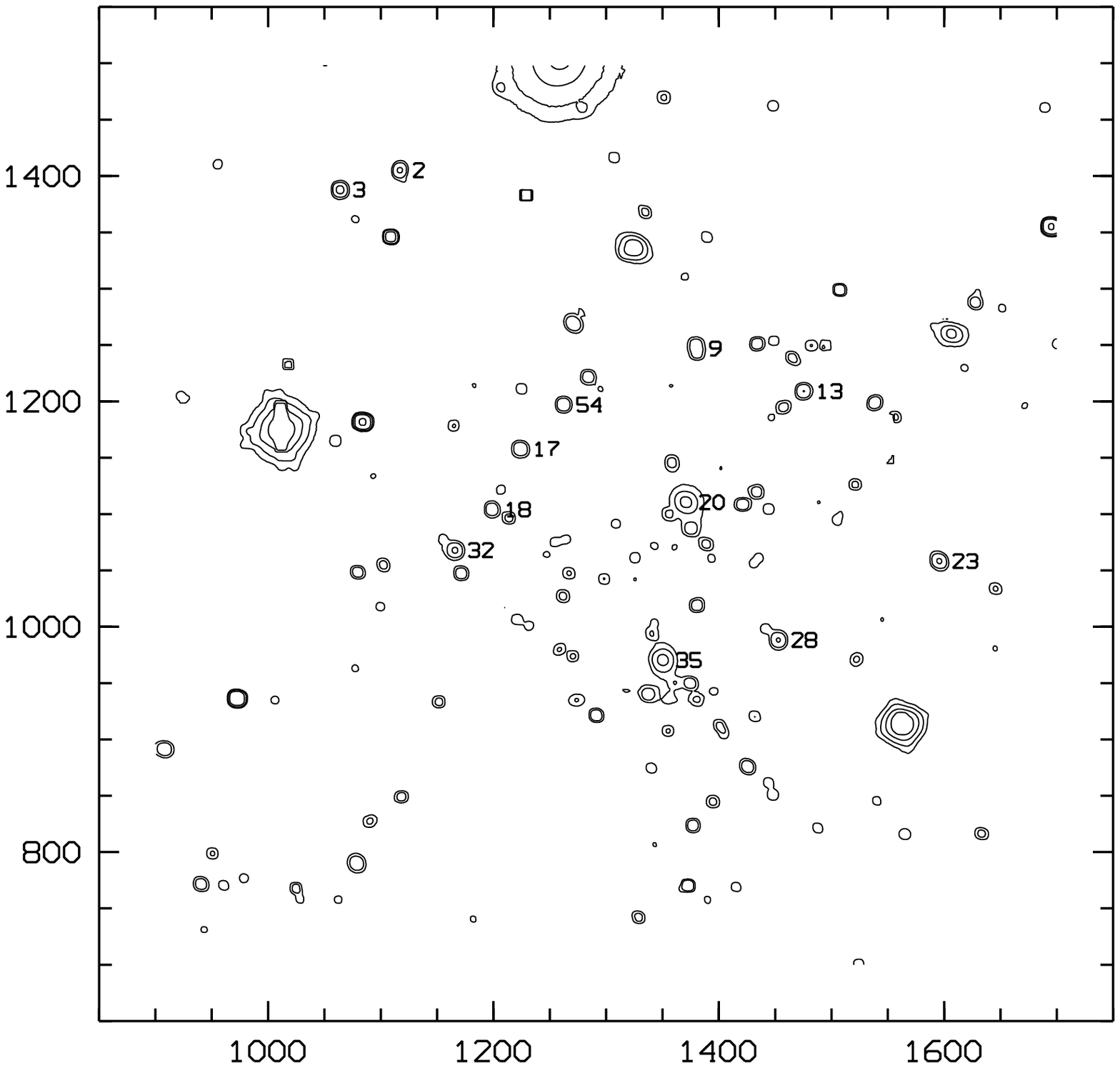,width=175mm}
\caption{\abell\ at $z=0.375$. Coordinates are CCD pixels and correspond to
columns X and Y of Table~\ref{tab_a370}. Labels follow column ID.}
\label{fig_a370}
\end{figure*}
%

%
\begin{table*}
\centering
\begin{minipage}{105mm}
\caption{Photometric data of galaxies with spectra in \abell.}
\label{tab_a370}
\begin{tabular}{rcrccccccc}
\multicolumn{1}{c}{ID} & \multicolumn{1}{c}{BOW} &
\multicolumn{1}{c}{PK} & X & Y & V & R & I & V--R & V--I  \\
\hline
A02 & 31 & 132 & 1116.8 & 1405.2 & 20.01 & 18.93 & 18.17 & 1.08 & 1.85 \\
A03 & 22 &     & 1063.7 & 1387.9 & 20.22 & 19.09 & 18.28 & 1.13 & 1.94 \\
A13 & 34 & 107 & 1475.2 & 1209.2 & 20.42 & 19.19 & 18.41 & 1.23 & 2.01 \\
A17 & 26 &  97 & 1223.9 & 1157.9 & 20.40 & 19.19 & 18.40 & 1.21 & 2.00 \\
A18 & 41 &  88 & 1198.6 & 1104.2 & 20.55 & 19.42 & 18.63 & 1.13 & 1.90 \\
A20 & 10 &  90 & 1370.9 & 1110.6 & 19.77 & 18.56 & 17.76 & 1.21 & 2.01 \\
A23 & 21 &  70 & 1595.6 & 1058.3 & 20.18 & 19.04 & 18.22 & 1.14 & 1.96 \\
A28 & 29 &  53 & 1452.8 &  988.3 & 20.13 & 19.06 & 18.25 & 1.07 & 1.88 \\
A32 &    &  76 & 1165.7 & 1067.9 & 20.13 & 19.02 & 18.20 & 1.11 & 1.93 \\
\end{tabular}

\medskip
ID numbers correspond to the reference number of Table~1 in \cite{MSFM88}, 
BOW is the reference number of Table~3 in \cite{BOW83}, while
PK is the reference number of Table~4b in \cite{PK91}.
\end{minipage}
\end{table*}
\begin{figure*}
\psfig{figure=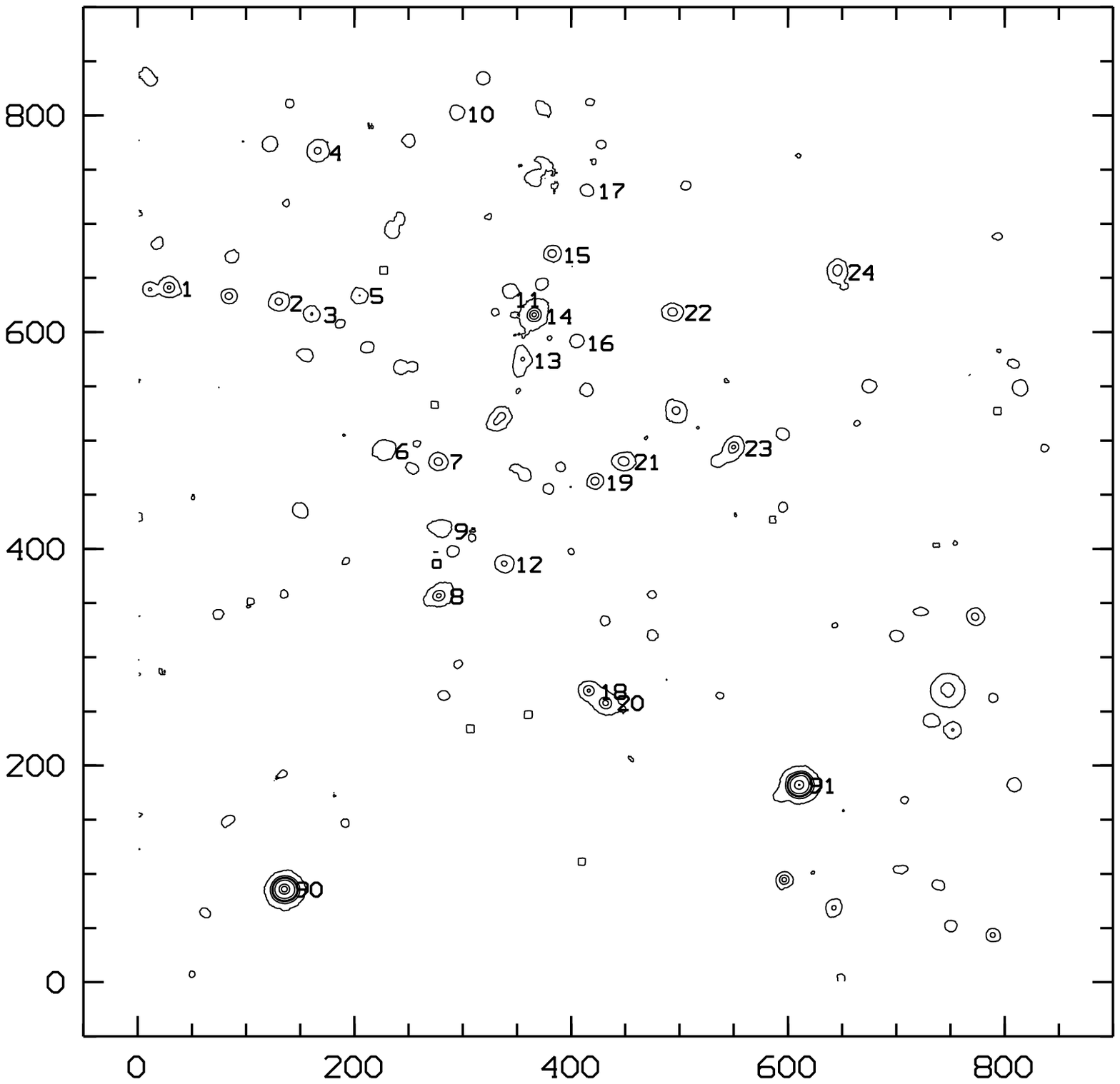,width=175mm}
\caption{\cl\ at $z=0.377$. Coordinates are CCD pixels and correspond to
columns X and Y of Table~\ref{tab_cl0949}. Labels follow column ID.}
\label{fig_cl0949}
\end{figure*}
%

%
\begin{table*}
\centering
\begin{minipage}{95mm}
\caption{Photometric data of galaxies with spectra in \cl}
\label{tab_cl0949}
\begin{tabular}{rrcccccrr}
\multicolumn{1}{c}{ID} & \multicolumn{1}{c}{DG} & X & Y & V & R & I & 
\multicolumn{1}{c}{V--R} & \multicolumn{1}{c}{V--I} \\
\hline
C04 &  80 & 164.8 & 765.5 & 20.95 & 19.90 & 18.60 &  1.05 &  2.35 \\
C08 & 193 & 276.7 & 356.2 & 20.81 & 19.43 & 18.71 &  1.38 &  2.10 \\
C14 & 118 & 364.4 & 614.6 & 20.52 & 19.20 & 17.99 &  1.32 &  2.53 \\
C18 & 217 & 414.7 & 268.1 & 21.16 & 20.31 & 18.83 &  0.85 &  2.33 \\
C20 & 221 & 430.5 & 257.6 & 21.74 & 19.81 & 18.47 &  1.93 &  3.27 \\
C23 &     & 548.2 & 492.7 & 20.59 & 19.67 & 18.43 &  0.92 &  2.16 \\
\end{tabular}

\medskip
DG is the reference number of Table~2 in \cite{DG92}.
\end{minipage}
\end{table*}
\begin{figure*}
\psfig{figure=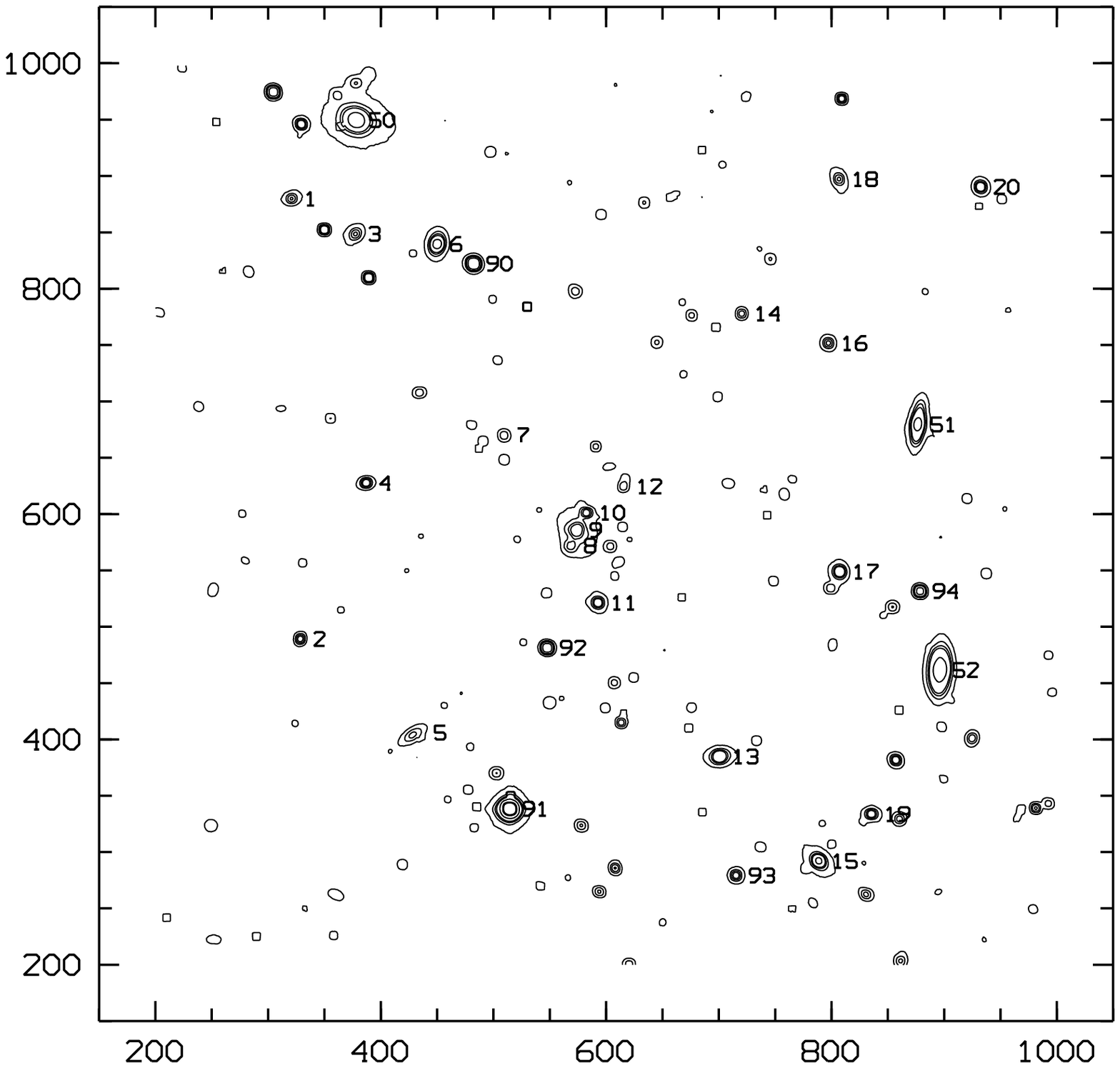,width=175mm}
\caption{\ms\ at $z=0.372$. Coordinates are CCD pixels and correspond to
columns X and Y of Table~\ref{tab_ms1512}. Labels follow column ID.}
\label{fig_ms1512}
\end{figure*}
\begin{table}
\caption{Photometric data of galaxies with spectra in \ms}
\label{tab_ms1512}
\begin{tabular}{rcccccrc}
\multicolumn{1}{c}{ID} & X & Y & V & R & I & \multicolumn{1}{c}{V--R} & 
V--I \\
\hline
M02 & 328.6 & 489.2 & 21.74 & 20.40 & 19.40 & 1.34 & 2.34 \\
M09 & 574.0 & 585.7 & 21.02 & 19.79 & 18.70 & 1.23 & 2.32 \\
M11 & 592.8 & 521.4 & 21.38 & 20.16 & 18.97 & 1.22 & 2.41 \\
M15 & 788.7 & 292.3 & 20.47 & 19.31 & 18.29 & 1.16 & 2.18 \\
M17 & 807.3 & 548.9 & 20.94 & 19.82 & 18.77 & 1.12 & 2.17 \\
M19 & 835.5 & 333.9 & 21.49 & 20.23 & 19.31 & 1.26 & 2.18 \\
\end{tabular}
\end{table}

\section{Spectroscopic Data}

%
\begin{table*}
\centering
\begin{minipage}{135mm}
\caption{Spectroscopic data}
\label{tab_spek} 
\begin{tabular}{cccccccccccc}
ID & $v_r$ & $\sigma$ & $\Delta\sigma$ & $\lg\sigma$ & Mg$_b$ &
Mg$_{b,\rmn cor}$ & $\Delta{\rmn Mg}_b$ & $\Delta{\rmn Mg}_{b,\rmn evo}$ &
H$_\beta$ & $\Delta{\rmn H}_\beta$ & H$_{\beta,\rmn cor}$ \\
\hline
A02 & 108994 & 270 & 30 & 2.47 & 3.91 & 5.04 & 0.23 & 0.01   & 1.18 & 0.19 &  1.02 \\
A03 & 108019 & 270 & 20 & 2.47 & 3.78 & 4.89 & 0.19 & --0.14 & 2.75 & 0.16 &  2.65 \\
A13 & 113224 & 250 & 30 & 2.44 & 3.73 & 4.75 & 0.36 & --0.18 & 0.88 & 0.28 &  0.71 \\
A17 & 114537 & 215 & 20 & 2.37 & 3.38 & 4.27 & 0.25 & --0.49 & 1.68 & 0.20 &  1.52 \\
A18 & 112877 & 230 & 80 & 2.40 & 3.51 & 4.46 & 0.25 & --0.38 & 1.69 & 0.21 &  1.53 \\
A20 & 113350 & 300 & 40 & 2.52 & 3.45 & 4.62 & 0.17 & --0.53 & 0.85 & 0.14 &  0.68 \\
A23 & 110390 & 280 & 20 & 2.49 & 3.75 & 4.89 & 0.23 & --0.18 & 1.94 & 0.19 &  1.81 \\
A28 & 111275 & 245 & 20 & 2.43 & 3.80 & 4.82 & 0.23 & --0.10 & 1.60 & 0.19 &  1.45 \\
A32 & 110591 & 185 & 20 & 2.31 & 3.18 & 3.98 & 0.18 & --0.60 & 1.52 & 0.15 &  1.35 \\ 
C04 & 104170 & 120 & 30 & 2.12 & 2.26 & 2.92 & 0.40 & --1.15 & 1.86 & 0.33 &  1.68 \\
C08 & 113600 & 230 & 25 & 2.40 & 2.86 & 3.74 & 0.21 & --1.10 & 2.20 & 0.17 &  2.05 \\
C14 & 114560 & 320 & 50 & 2.55 & 3.81 & 5.13 & 0.22 & --0.10 & 1.83 & 0.18 &  1.71 \\ 
C18 & 112860 & 125 & 15 & 2.14 & 3.05 & 3.74 & 0.22 & --0.39 & 1.78 & 0.18 &  1.60 \\ 
C20 & 113090 & 230 & 25 & 2.40 & 3.61 & 4.56 & 0.17 & --0.28 & 1.72 & 0.14 &  1.57 \\
C23 & 114010 & 225 & 25 & 2.39 & 2.42 & 3.25 & 0.25 & --1.57 & 1.73 & 0.20 &  1.58 \\
M02 & 111655 & 200 & 20 & 2.34 & 3.76 & 4.64 & 0.16 & --0.04 & 1.99 & 0.18 &  1.83 \\ 
M09 & 111514 & 260 & 20 & 2.46 & 3.63 & 4.68 & 0.12 & --0.30 & 0.85 & 0.08 & -1.07 \\
M11 & 111807 & 160 & 20 & 2.25 & 3.63 & 4.41 & 0.40 & --0.01 & 1.66 & 0.10 &  1.49 \\ 
M15 & 111868 & 285 & 20 & 2.50 & 3.30 & 4.40 & 0.13 & --0.70 & 1.82 & 0.09 &  1.69 \\
M17 & 109135 & 200 & 15 & 2.34 & 2.86 & 3.67 & 0.30 & --1.01 & 1.90 & 0.09 &  1.74 \\ 
M19 & 110515 & 215 & 15 & 2.37 & 3.38 & 4.27 & 0.28 & --0.49 & 1.81 & 0.24 &  1.66 \\
\end{tabular}

\medskip
ID refers to Tables~\ref{tab_a370}, \ref{tab_cl0949} and
\ref{tab_ms1512},
$v_r$ is the measured radial velocity in km s$^{-1}$ (with an average error 
of ca. $\pm 20$\,km s$^{-1}$),
$\sigma$ the measured \veldis\ in km s$^{-1}$,
$\Delta\sigma$ the error thereof,
$\lg\sigma$ the decimal logarithm of the aperture corrected \veldis,
Mg$_b$ the measured \mgl\ in \AA,
Mg$_{b,\rmn cor}$ the \mgl\ corrected for \veldis\ and aperture,
$\Delta{\rmn Mg}_b$ the error thereof,
$\Delta{\rmn Mg}_{b,\rmn evo}$ the evolution of \mgl\ between $z=0$ and 
$z=0.37$,
\hb\ the measured \hb\ \lst\ in \AA,
$\Delta{\rmn H}_\beta$ the error thereof and
H$_{\beta,\rmn cor}$ the \hb\ \lst\ corrected for \veldis.
\end{minipage}
\end{table*}
%

%
\begin{table*}
\centering
\begin{minipage}{95mm}
\caption{Absolute $B$--magnitudes 
($H_0=50{\rmn \:km\:s^{-1}\:Mpc^{-1}}, q_0=0.5.$)}
\label{tab_spek2} 
\begin{tabular}{ccccccc}
ID & V$_{\rmn tot}$ & $\Delta\mbox{V}$ & M$_B$ & 
$\Delta\mbox{M}_{B\rmn ,evo}$
& M$_{B\rm ,cor}$ & $\Delta\mbox{M}_{B\rmn ,cor}$ \\
\hline
A02 & 19.35 & 0.26 &  --22.79 & --0.01 & --22.80 & 0.36 \\
A03 & 19.56 & 0.26 &  --22.58 &  0.19 & --22.39 & 0.33 \\
A13 & 19.90 & 0.26 &  --22.25 &  0.26 & --21.99 & 0.49 \\
A17 & 20.09 & 0.26 &  --22.05 &  0.68 & --21.37 & 0.38 \\
A18 & 20.16 & 0.26 &  --21.99 &  0.53 & --21.45 & 0.38 \\
A20 & 18.90 & 0.26 &  --23.25 &  0.74 & --22.50 & 0.32 \\
A23 & 19.45 & 0.26 &  --22.69 &  0.25 & --22.44 & 0.37 \\
A28 & 19.64 & 0.26 &  --22.50 &  0.13 & --22.37 & 0.37 \\
A32 & 19.96 & 0.26 &  --22.18 &  0.84 & --21.33 & 0.33 \\
C04 & 20.93 & 0.27 &  --21.21 &  1.61 & --19.60 & 0.54 \\
C08 & 20.34 & 0.27 &  --21.80 &  1.54 & --20.26 & 0.36 \\
C14 & 19.37 & 0.27 &  --22.78 &  0.13 & --22.64 & 0.37 \\
C18 & 21.14 & 0.27 &  --21.01 &  0.54 & --20.46 & 0.37 \\
C20 & 21.27 & 0.27 &  --20.87 &  0.39 & --20.48 & 0.34 \\
C23 & 20.15 & 0.27 &  --21.99 &  2.19 & --19.80 & 0.40 \\
M02 & 20.95 & 0.27 &  --21.20 &  0.05 & --21.14 & 0.32 \\
M09 & 19.77 & 0.27 &  --22.38 &  0.42 & --21.95 & 0.30 \\
M11 & 21.03 & 0.27 &  --21.11 &  0.01 & --21.10 & 0.55 \\
M15 & 18.90 & 0.27 &  --23.24 &  0.98 & --22.26 & 0.31 \\
M17 & 20.25 & 0.27 &  --21.89 &  1.40 & --20.49 & 0.44 \\
M19 & 20.53 & 0.27 &  --21.61 &  0.68 & --20.93 & 0.43 \\
\end{tabular}
      
\medskip
ID refers to Tables~\ref{tab_a370}, \ref{tab_cl0949} and
\ref{tab_ms1512},
V$_{\rm tot}$ is the total apparent $V$--magnitude,
$\Delta\mbox{V}$ the error thereof,
M$_B$ the absolute $B$--magnitude in the restframe of the galaxy,
$\Delta\mbox{M}_{B\rm .evo}$ the evolutionary correction according to 
equation~\ref{gl_mgbmb},
M$_{B\rm ,cor}$ the absolute $B$--magnitude corrected for evolution and 
$\Delta\mbox{M}_{B\rm ,cor}$ the error thereof.
\end{minipage}
\end{table*}
%

Spectra and tables with data relevant for the \mgsr\ and \fjr\ are given.
All spectra were taken with the 3.5m-telescope on Calar Alto with the TWIN
spectrograph and cover the wavelength range $\lambda\lambda=6500-7500$\,\AA.
The instrumental resolution was 105~km~s$^{-1}$. The absorption lines of
\hb\ (\lo\ 4861\,\AA), \mg\ ($\lambda_0 \approx$ 5173\,\AA), Fe5270 (\lo\
5269\,\AA) and Fe5335 (\lo\ 5328\,\AA) are readily visible as well as the
telluric B--band ($\lambda_0 \approx 6900$\,\AA).

\begin{figure*}
\psfig{figure=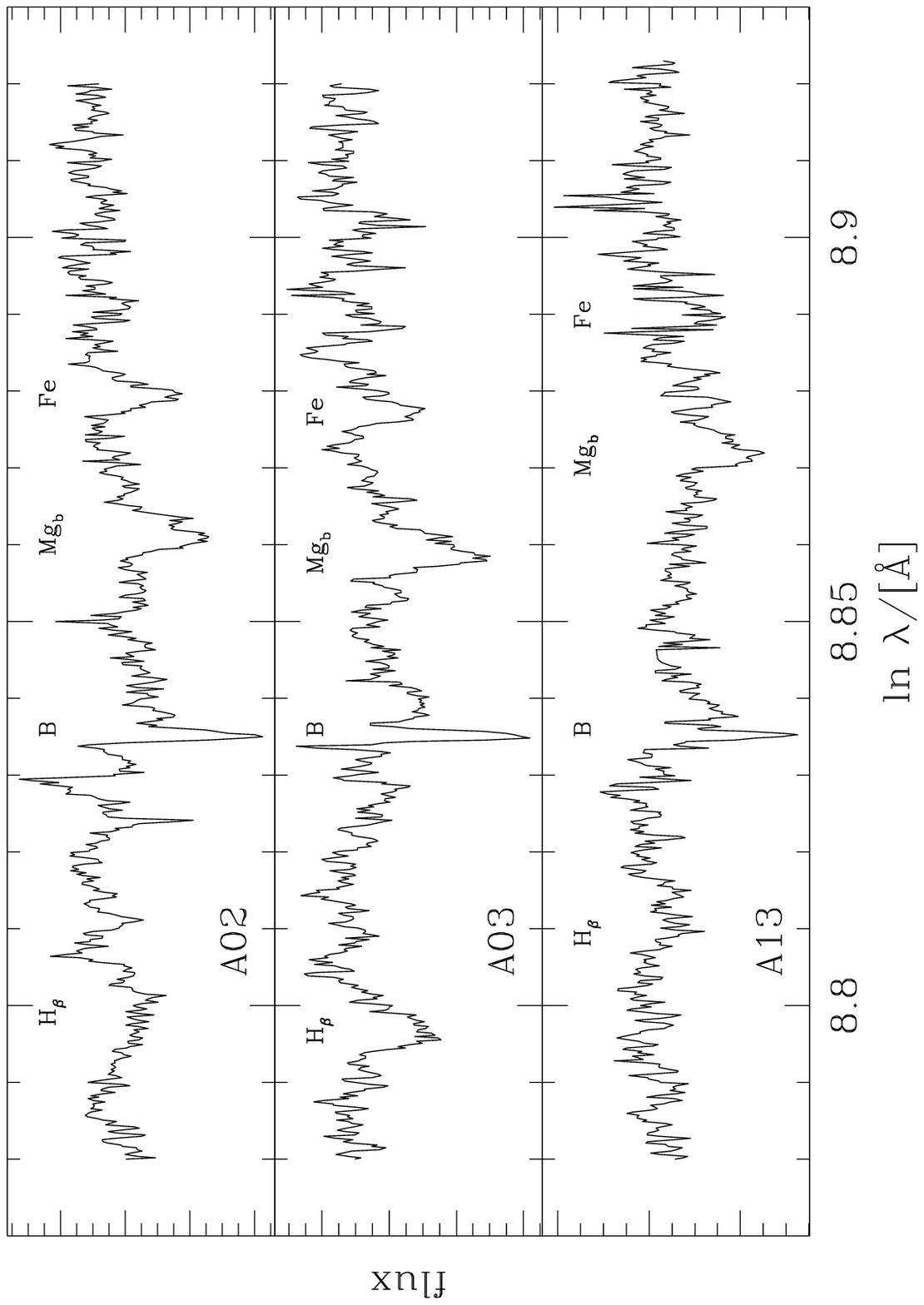,width=175mm}
\caption{Spectra of galaxies A02, A03 and A13}
\end{figure*}
\begin{figure*}
\psfig{figure=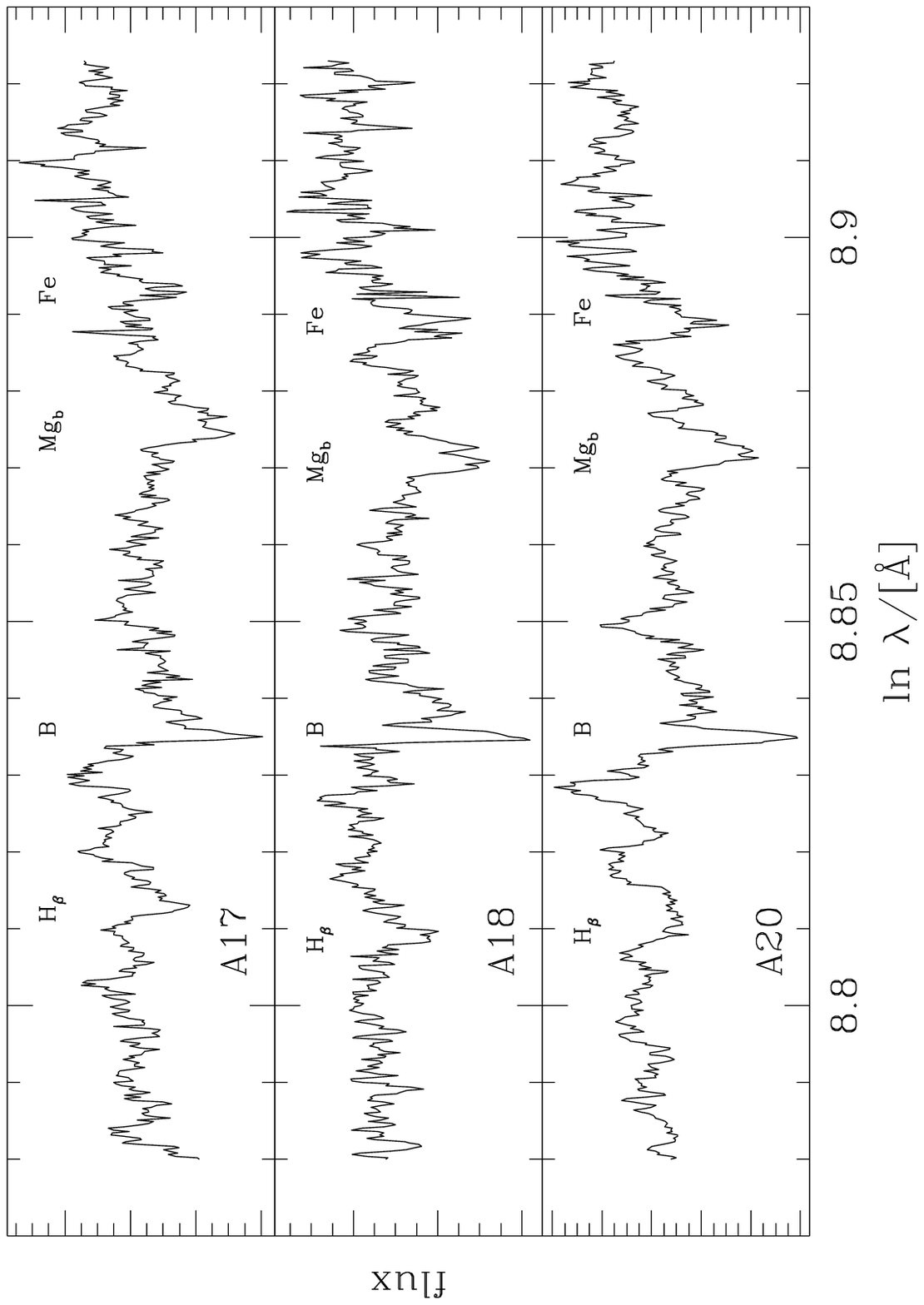,width=175mm}
\caption{Spectra of galaxies A17, A18 and A20}
\end{figure*}
\begin{figure*}
\psfig{figure=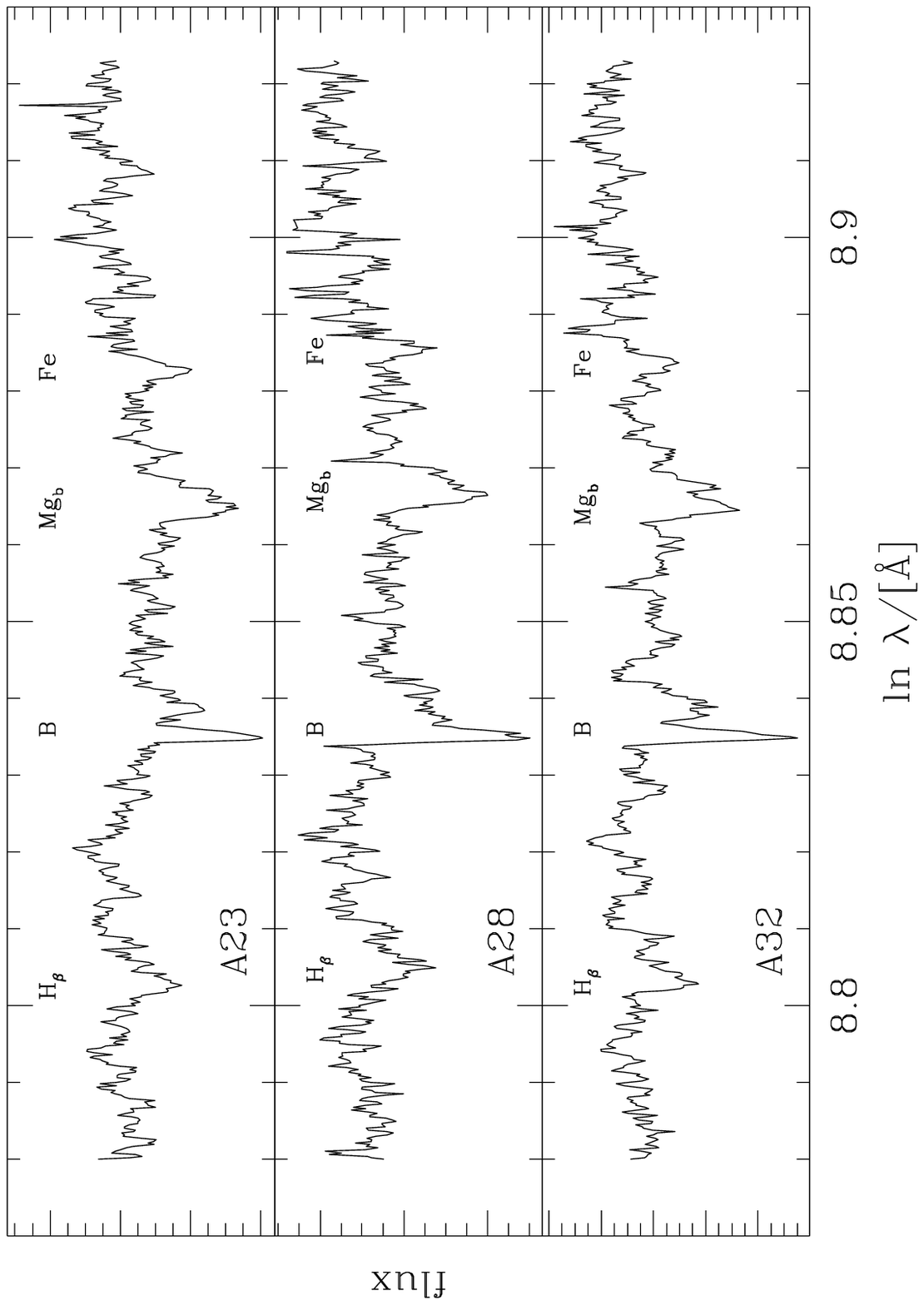,width=175mm}
\caption{Spectra of galaxies A23, A28 and A32}
\end{figure*}
\begin{figure*}
\psfig{figure=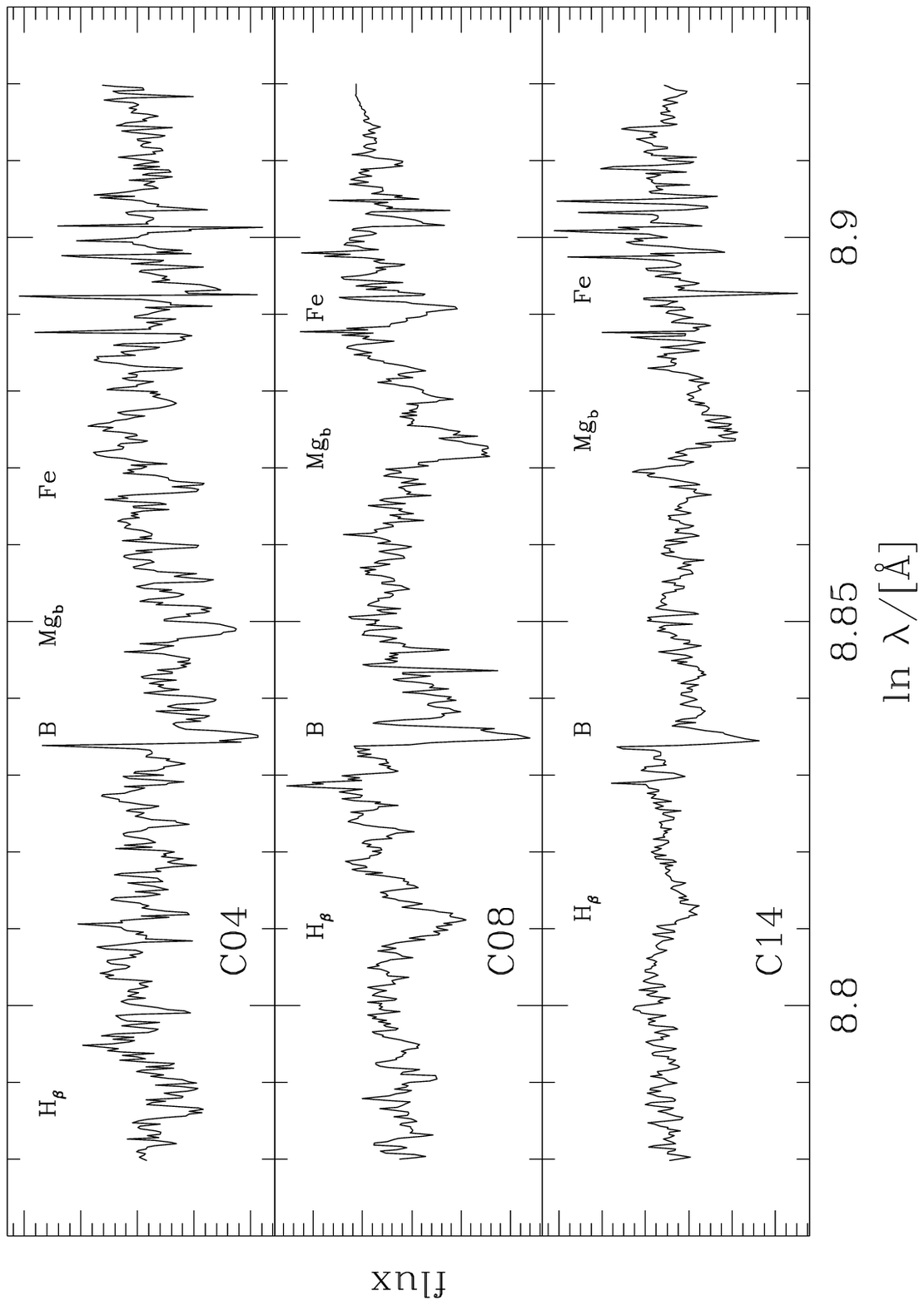,width=175mm}
\caption{Spectra of galaxies C08, C18 and C20}
\end{figure*}
\begin{figure*}
\psfig{figure=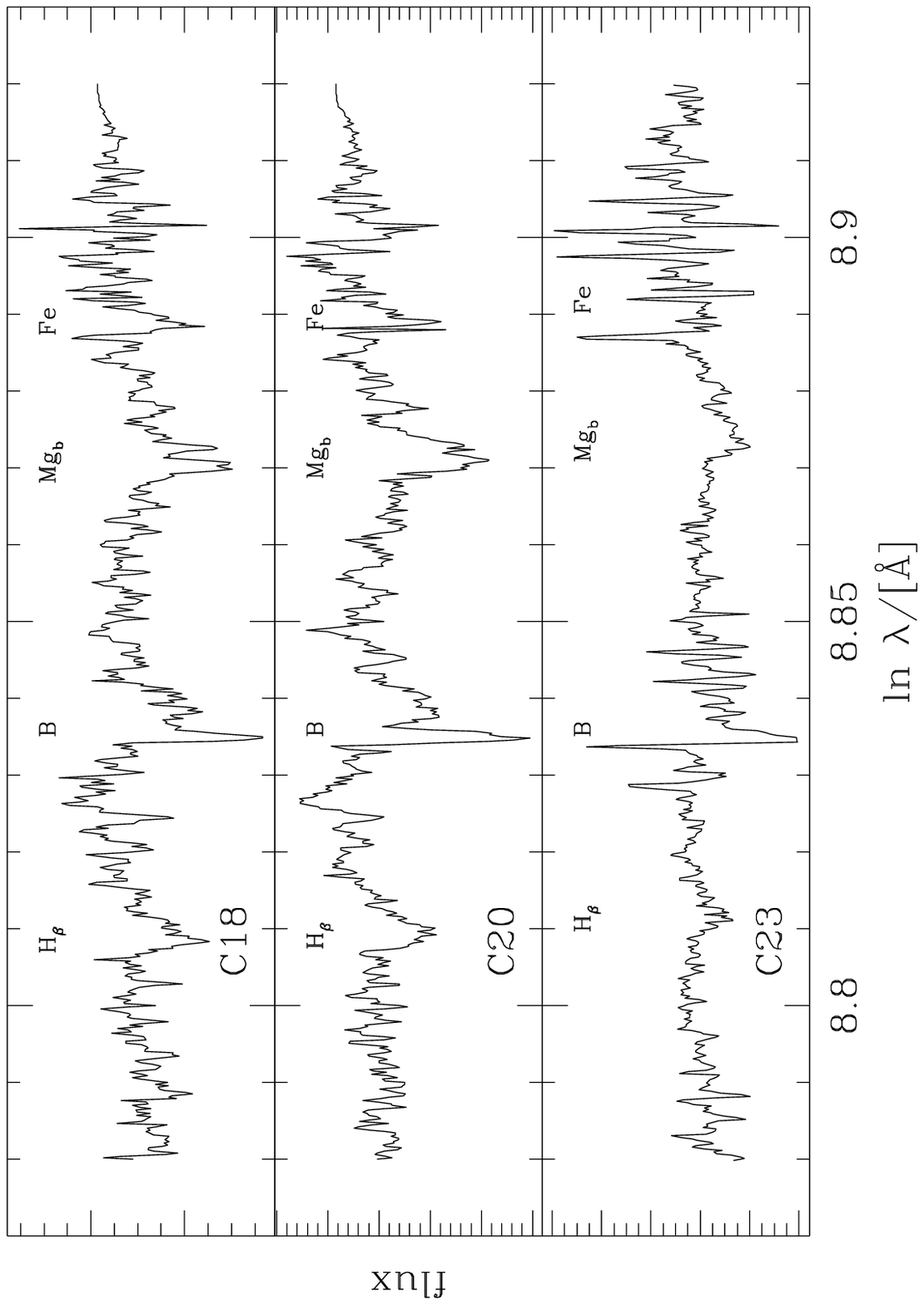,width=175mm}
\caption{Spectra of galaxies C04, C14 and C23}
\end{figure*}
\begin{figure*}
\psfig{figure=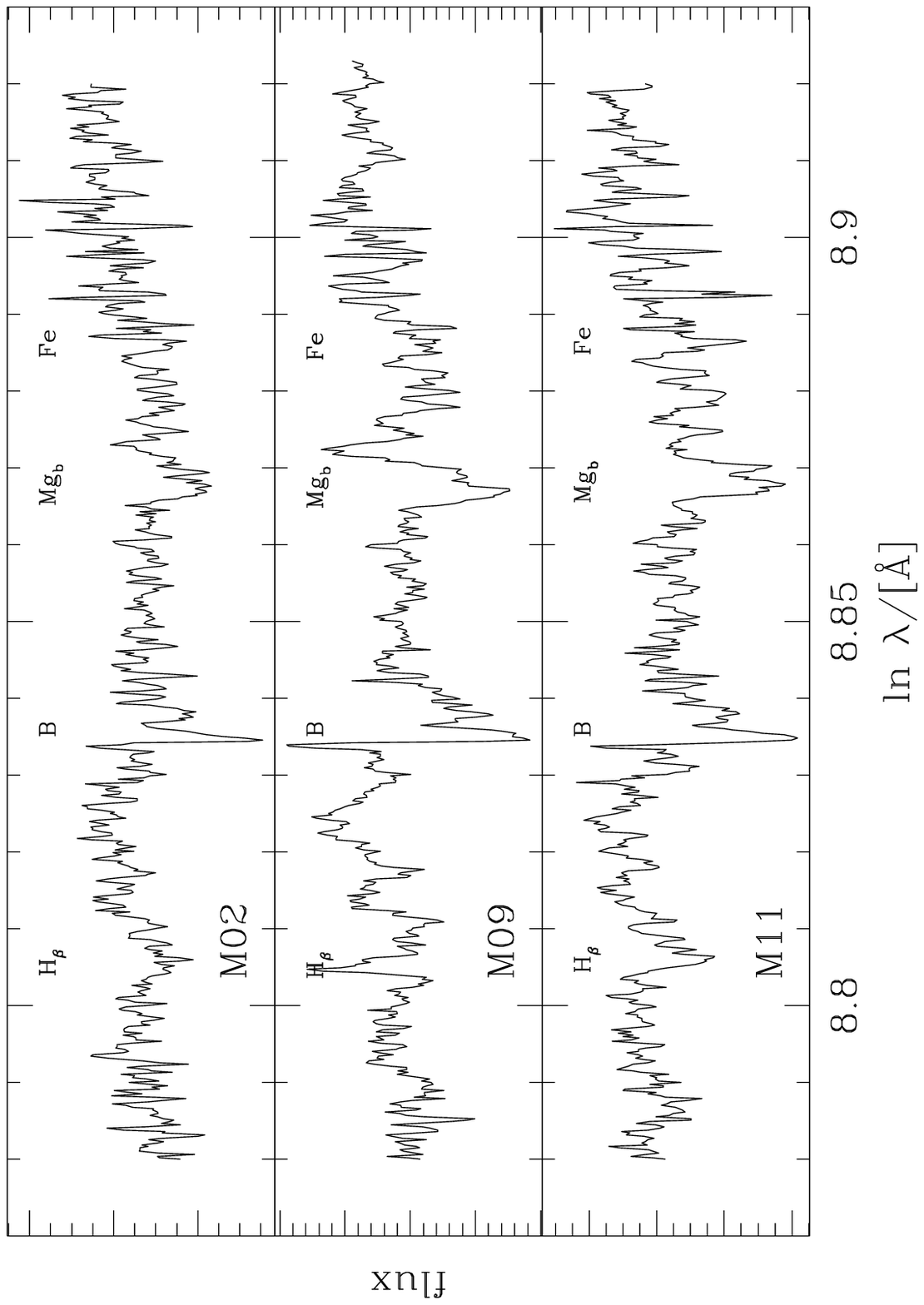,width=175mm}
\caption{Spectra of galaxies M02, M09 and M11}
\end{figure*}
\begin{figure*}
\psfig{figure=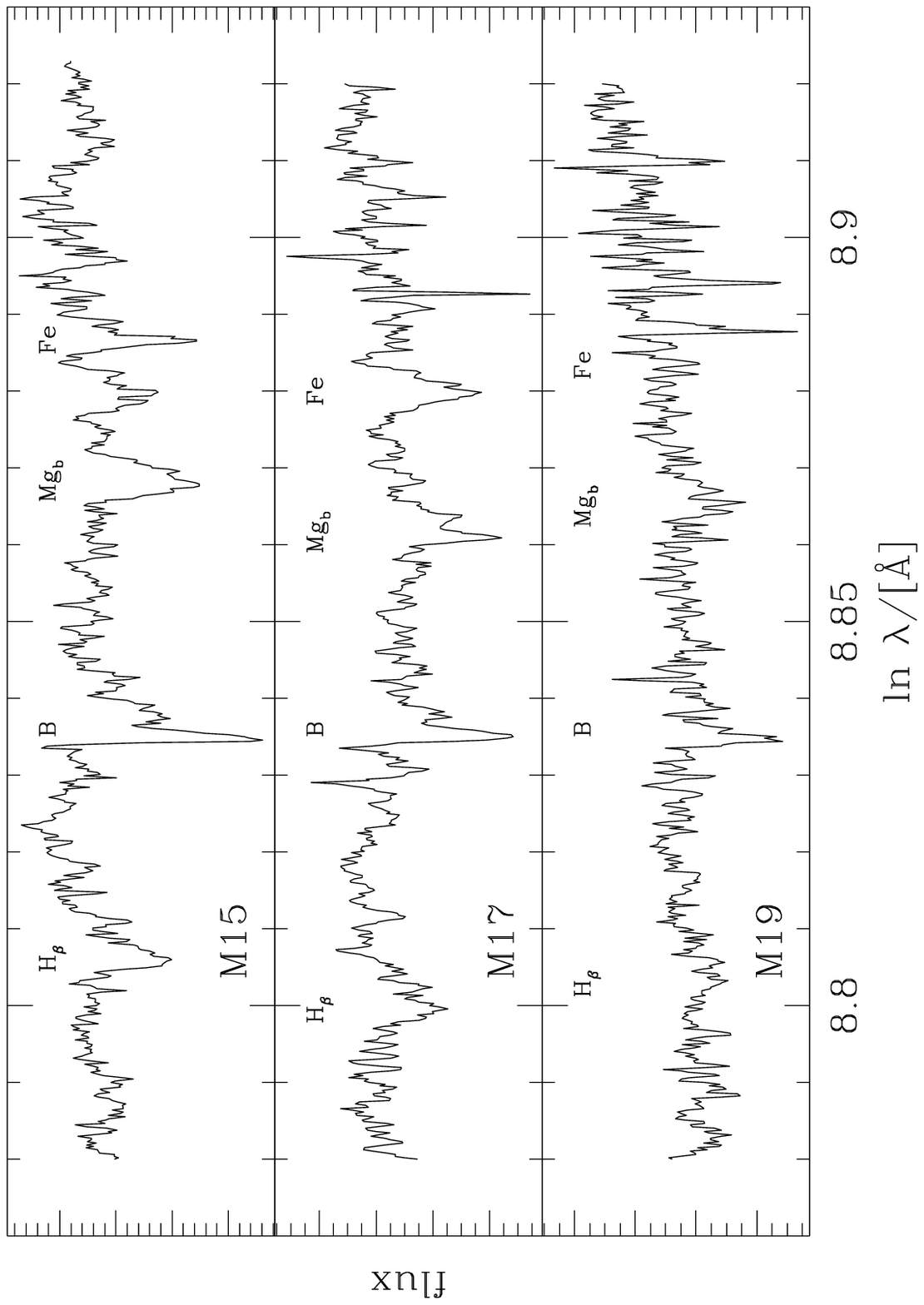,width=175mm}
\caption{Spectra of galaxies M15, M17 and M19}
\end{figure*}

\bsp

\label{lastpage}

\end{document}